\documentclass[conference, 10pt]{IEEEtran}
\IEEEoverridecommandlockouts
\usepackage{cite}
\usepackage{amsmath,amssymb,amsfonts}
\usepackage{graphicx}
\usepackage{textcomp}
\usepackage{xcolor}
\usepackage{subcaption}
\usepackage{bm}
\usepackage{multirow}
\usepackage{booktabs}
\usepackage{subcaption}
\usepackage{amsthm}\usepackage{balance}\usepackage{url}
\usepackage{algorithm,algpseudocode}
\usepackage{amssymb}
\usepackage{siunitx}

\renewcommand{\appendixname}{Appendix~}

\newtheorem*{proof*}{Proof}

\algrenewcommand\algorithmicrequire{\textbf{Initialization:}}
\algrenewcommand\algorithmicensure{\textbf{Output:}}
\algdef{SE}[DOWHILE]{Do}{doWhile}{\algorithmicdo}[1]{\algorithmicwhile\ #1}%

\makeatletter

\newcommand{\Rmnum}[1]{\expandafter\@slowromancap\romannumeral #1@}
\makeatother

\date{}

\def\BibTeX{{\rm B\kern-.05em{\sc i\kern-.025em b}\kern-.08em
    T\kern-.1667em\lower.7ex\hbox{E}\kern-.125emX}}

\begin{document}
\title{GenEditSC:  Generative Editing Semantic Communication with Multimodal Image Editing}

\author{Shuoyao Wang~\IEEEmembership{Senior Member,~IEEE},  Suzhi Bi~\IEEEmembership{Senior Member,~IEEE},  Mingze Gong, Zhanpeng Wang,  
 \\  Li Ping Qian~\IEEEmembership{Senior Member,~IEEE}, and Qiang Ye~\IEEEmembership{Senior Member,~IEEE}
\thanks{S. Wang, S. Bi, Z. Wang, and M. Gong are with College of Electronics and Information Engineering, Shenzhen University, China. L. Qian is School of Ocean Engineering and Technology, Sun Yat-sen University, Zhuhai, China. Q. Ye  is with the Department of Electrical and Software Engineering,  University of Calgary, Canada.}
}
\maketitle

\begin{abstract}\label{abstract}
    Deep learning-based joint source-channel coding  has recently demonstrated strong potential for semantic communication (SemComm).
    However, most existing approaches focus on optimizing visual-fidelity metrics, which can lead to reduced perceptual quality. 
    Generative model-based SemComm leverages rich prior knowledge from large-scale pre-training to enhance perceptual quality, but often at the cost of increased structural distortion.
    This paper addresses  {the} above issues by proposing a two-stage semantic image transmission framework, integratinga fixed and replaceable multimodal image editing model at the receiver.
    In the first stage, {a JSCC-based discriminative transmission selectively prioritizes semantically important regions, preserving scene layout and object integrity under limited bandwidth.}
   {In} the second phase, generative image editing refines missing or degraded details based on the textual descriptions, enhancing semantic fidelity and perceptual quality.
     {Extensive experiments show that GenEditSC outperforms the evaluated discriminative and generative baselines in semantic preservation, perceptual quality, and visual fidelity across various channel conditions and compression ratios.}
\end{abstract}
\begin{IEEEkeywords}
    Semantic communication, joint source-channel coding, generative image editing, image transmission.
\end{IEEEkeywords}


\section{Introduction}\label{introduction}
    The sixth generation (6G) is envisioned as an intelligent information system, both powered by and driving advances in deep learning (DL) technology.
    Leveraging DL breakthroughs, semantic communication (SemComm) systems typically employ deep neural networks (DNNs) for joint source-channel coding (JSCC), enabling efficient extraction, encoding, and transmission of semantic information across diverse data types, including images and text.
    In particular, the representative DeepJSCC~\cite{deepjscc} jointly trains neural encoders and decoders to learn robust feature mappings integrating source and channel coding, thereby achieving graceful performance degradation with SNR and outperforming conventional schemes.
    
    Despite the superior performance of current DL-based JSCC systems in semantic image transmission, they are typically optimized for pixel-level fidelity between the source and reconstructed images, often measured by metrics such as peak signal-to-noise ratio (PSNR).
    However, it is increasingly acknowledged that high pixel-level similarity does not necessarily translate into high perceptual quality, which better captures the realism and visual appeal as perceived by human observers \cite{ye2025low}.

    To enhance the perceptual quality of transmitted images, generative AI (GenAI) models have recently achieved remarkable success in enabling communication at the semantic level.
    Rather than transmitting dense pixel-level features, these approaches exploited generative models (e.g., Stable Diffusion~\cite{stable_diffusion}) to encode and send compact, high-level cues, such as image captions~\cite{jihongpark}, object layouts~\cite{mask}, or semantic segmentation maps~\cite{segmentation}, that emphasize semantically critical content while discarding redundant visual details.
    At the receiver, a generative model synthesizes the final image conditioned on these cues, {i.e. conditional generation,} enabling more efficient bandwidth utilization and greater flexibility in adapting to task-specific requirements.
    Leveraging rich prior knowledge from large-scale pre-training, generative SemComm often produces results that are perceptually rich.
    
    
    Despite these advances, generative SemComm still faces two main challenges:
    i) The output quality is highly sensitive to the precision of the transmitted conditions. Using fine-grained cues, such as segmentation maps~\cite{diff-go} and latent embeddings~\cite{Ke2025resulic}, substantially increases communication overhead. Conversely, relying on coarse-grained semantic conditions (e.g., class labels or captions) fails to capture sufficient visual details, leading to noticeable mismatches in spatial layout, color fidelity, and textural consistency between the reconstructed and original images~\cite{jihongpark}. 
    ii) 
    Due to the cross-modal predictive nature,  generative rendering suffers from inherent randomness in rendering outcomes~\cite{Conquer}.


    Overall, achieving high-performance generative SemComm requires addressing two fundamental open questions:
    \begin{itemize}
        \item What semantic conditions should be transmitted to balance communication efficiency and generation quality?
        \item  {How should the JSCC front-end be designed and trained to produce an intermediate reconstruction that better supports a fixed generative editing backend, without jointly optimizing the backend?}
    \end{itemize}
    
    For Q1, recent advances in pretrained multimodal generative image models have enabled high-quality text-guided image generation and editing\footnote{ {\url{https://github.com/QwenLM/Qwen-Image}}}. 
    Inspired by this, we propose Generative Editing Semantic Communication (GenEditSC), a novel paradigm that enables efficient semantic image transmission and controllable content refinement.
    \textit{Unlike prior Generative SemComm approaches that mainly focus on generative rendering (i.e., cross-modal generation), we are among the first to explore a generative editing framework that integrates discriminative and generative processes.} This editing-based approach offers distinct advantages by preserving the structural fidelity of the original input while allowing for targeted, semantically-consistent enhancements, thereby mitigating the risk of entire hallucination or structural distortion often associated with full generative rendering.

   For Q2, we argue that achieving higher pixel fidelity via conventional JSCC does not necessarily lead to better perceptual quality,  {as it often wastes bandwidth on semantically redundant regions (e.g., background). 
    To this end, we develop a two-stage semantic importance-aware framework. { Here, ``semantic importance'' refers to the spatial concentration of semantically important information within an image, quantified through feature-level importance scores derived from multimodal semantic alignment.} In Stage I, a discriminative JSCC encoder performs semantically-aware compression. It prioritizes the transmission of crucial foreground details while aggressively compressing less important background regions, significantly reducing data transmission under limited bandwidth without sacrificing structural integrity. 
     {In Stage II, a  {fixed} generative editing backend synthesizes missing or degraded details from the accurately transmitted layout and semantic context, thereby improving perceptual quality while preserving the source structure.}
    }

  {Overall, this hybrid discriminative-generative paradigm achieves a superior trade-off among semantic fidelity, perceptual quality, and visual fidelity compared to conventional approaches.} The main contributions of this work are summarized as follows:
    \begin{itemize}
   \item        \textit{A hybrid generative editing semantic communication framework (GenEditSC):}
     {We propose a hybrid discriminative-generative semantic communication framework that integrates a semantic-aware JSCC transmission front-end with a  {fixed} and replaceable generative editing backend. By adopting a generative editing paradigm rather than direct cross-modal rendering, the proposed framework preserves source structure while enabling controllable semantic refinement, thereby improving the trade-off among transmission efficiency, semantic fidelity, and perceptual quality.}


        \item { To address the unequal importance of different image regions in downstream generative editing, we propose a  {foreground/background dual-branch transmission architecture} that uses an asymmetric dual-branch architecture with different functional roles for foreground and background processing.  {To enable this decomposition, we further introduce a semantic importance prediction module for explicit semantic-pixel alignment, which guides the separation of transmitted representations into foreground-detail and background-context branches.}
        }
        
 \item{ {To avoid the computational cost of jointly fine-tuning the communication modules and the large generative backend, we introduce a semantic-importance-aware JSCC objective that optimizes the trainable transmission front-end for downstream generative editing rather than conventional pixel-level reconstruction alone. The proposed loss function jointly considers pixel-semantic matching, foreground-detail reconstruction, and background semantic preservation, enabling a favorable trade-off among semantic fidelity, perceptual quality, and visual realism.}}

        \item  {Extensive experiments demonstrate that GenEditSC outperforms the evaluated discriminative JSCC and generative semantic communication baselines under matched settings in semantic fidelity, perceptual quality, and bandwidth efficiency. The results verify superior semantic
        	fidelity, perceptual quality, and bandwidth efficiencys. In addition, a 2-bit-quantized Qwen-Image backend is deployed on an NVIDIA Jetson AGX Orin 64~GB to evaluate the edge deployment feasibility.}
    \end{itemize}

    The remainder of this paper is organized as follows: Section \uppercase\expandafter{\romannumeral2} reviews related work on discriminative, generative, and LLM-edited semantic communication. 
    Section \uppercase\expandafter{\romannumeral3} introduces the overall system model of the proposed framework. 
    Section \uppercase\expandafter{\romannumeral4} details the network architecture. 
      {Section \uppercase\expandafter{\romannumeral5} presents quantitative, communication-overhead, edge-deployment, qualitative, and ablation evaluations.} 
    Finally, Section \uppercase\expandafter{\romannumeral6} concludes the paper and discusses future research directions.

\section{Related Work}
    In this section, we first review the evolution of DeepJSCC-based SemComm approaches.
    We then focus on generative SemComm. 
    Finally, we discuss LLM-aided approaches, including both discriminative JSCC and generative SemComm.

\subsection{Discriminative SemComm}
    As an emerging communication paradigm, SemComm aims to enhance communication efficiency through artificial intelligence technologies
    DeepJSCC\cite{deepjscc} pioneered this paradigm by jointly learning source and channel coding with DNN-based autoencoders. 

    Inspired by the great success of diffusion models, recent research  {has} further incorporated diffusion-based enhancement modules into the Deep learning-based JSCC approaches.  
    A straightforward approach is to preserve the DNN-based encoder-decoder structure while introducing diffusion-based denoising in the latent space, thereby leveraging the prior knowledge of generative models to improve robustness and noise resilience\cite{cddm,sgdjscc,semantic_prior_aided}. 
    However, discriminative decoders that focus primarily on visual fidelity still suffer from degraded perceptual quality, even when the latent features are refined.
    
\subsection{Generative SemComm}
    Generative SemComm primarily includes methods based on generative adversarial networks (GANs) and diffusion models.
    In these approaches, the reconstruction process at the receiver side employs generative models, which leverage rich prior knowledge from large-scale pre-training to enhance perceptual quality.
    In GAN-based SemComm, Han \emph{et al.}\cite{10096372} utilized GANs to achieve semantic image transmission with high compression ratios while considering privacy protection, although this approach required a complex training process. 
    However, similar to GAN-based methods in image processing (e.g., super-resolution, deblurring), the training of GAN-based SemComm models \cite{10096372, he2022robust} is unstable and prone to mode collapse, resulting in repetitive or highly similar generated samples that compromise image quality and diversity.

    In contrast, diffusion models provide significant advantages in terms of generation quality and training stability.
    For example, Park \emph{et al.}\cite{jihongpark} proposed transmitting only textual descriptions of the source image, which are then decoded into images via pretrained text-to-image (T2I)  {diffusion} models.
    While \cite{jihongpark} significantly reduces communication overhead, it often suffers from high stochasticity and poor structural consistency due to the inherent ambiguity of language prompts. 
    Accordingly, subsequent works incorporate auxiliary visual priors as conditions for generation: 
    i) Spatial layout cues such as coarse masks~\cite{mask} or edge maps~\cite{canny_latency} to localize main objects and maintain the overall structure; 
    ii) Semantic segmentation maps~\cite{diff-go} to preserve object boundaries and scene composition; 
    iii) Symbolic or relational priors, e.g. scene graphs~\cite{sg2sc} or agent-local maps~\cite{agent_driven_semcom}, to recover high-level structured remote scenes;
    iv)  Mixture of Semantics (MoS) scheme~\cite{MoS}, which seeks a balance by decomposing the image into regions of interest and non-interest for prioritized and lightweight transmission, respectively.


    
    {However, the above approaches share a critical trade-off: while richer condition signals, such as segmentation or depth maps, can boost reconstruction, they introduce substantial communication overhead, undermining the bandwidth efficiency that SemComm aims to achieve. 
    Conversely, lightweight conditions reduce transmission cost but fail to convey sufficient structural guidance, leading to semantic ambiguity and visual artifacts. 
    Thus, achieving high-fidelity, controllable image reconstruction under strict bandwidth constraints remains a key challenge.
    }

\subsection{LLM-based SemComm}
    In addition to adopting generative paradigms for image reconstruction, another trend in SemComm is the use of LLMs to enhance semantic fidelity, robustness, and adaptability. 
    Existing LLM-based SemComm studies can be broadly grouped into two categories.

\subsubsection{Semantic Importance Indicator}
    These works  {integrate} LLM reasoning into the source-channel coding process at the transmitter side,  to improve compression efficiency and resilience against channel impairments. 
    For instance, \cite{llm_semantic_power} uses a pre-trained LLM to  quantifie frame-level semantic importance for power allocation, while \cite{llm_enabled_semcom} integrates LLM tokenizers with unsupervised pre-training to construct a semantic knowledge base for improved decoding. 
    \cite{llm_end2end} treats LLM as a semantic codec, combining subword tokenization, adaptive rate control, and private-knowledge fine-tuning for robust transmission. 
    LaMoSC~\cite{lamosc} further positions LLM as a ``semantic hub'' that fuses visual--text cues for joint source--channel design, yielding better image quality at low SNR.  \cite{foundation_semcom} develops a LLM foundation-model-driven framework with perception-error-aware semantic power allocation, adapting transmission energy to cue importance for generative reconstruction.

\subsubsection{{Text Correction}}
    While the above methods improve efficiency and robustness,    they still suffer from the semantic inconsistency and reconstruction degradation at the receiver side caused by the noisy wireless channel.
    To address this, a growing body of work employs LLMs as \emph{semantic mediators}, not merely as compressors or classifiers, but as intelligent agents  that construct, refine, or repair textual prompts.    
  {For instance}, \cite{mlm_ood_semcom} leverages a LLM to refine textual descriptions via irrelevant-token filtering and contextual similarity, and \cite{mlm_privacy_semcom} enables privacy-preserving multimodal interpretation without exposing raw data. 
    Ref.~\cite{vlm_crossmodal_semcom} combines BLIP-based text extraction with Stable Diffusion, enhanced by BART text correction and continual learning for improved stability. 
    Prompt-based designs, such as \cite{crossmodal_prompt} and \cite{contextaware_prompt}, jointly optimize semantic encoding and context-aware prompt refinement, transmitting only filtered semantic cues for high-quality, low-bandwidth AIGC reconstruction. 
    The progressive scheme in \cite{progressive_text2img} transmits text incrementally, allowing LLM-based editing and completion when packets are lost before final image synthesis.

\subsubsection{{This work}}
    
Unlike prior works that primarily use LLMs for semantic understanding or textual refinement, GenEditSC employs a multimodal image editing model as a fixed receiver-side generative backend. The model takes both a spatially faithful JSCC reconstruction and a transmitted text prompt as conditions, thereby refining degraded details while retaining the source structure.

\section{System Model}\label{proposed_method}
\subsection{Overview}
\begin{figure*}[!t]   
	\centering        
	\includegraphics[clip,width=0.99\linewidth]{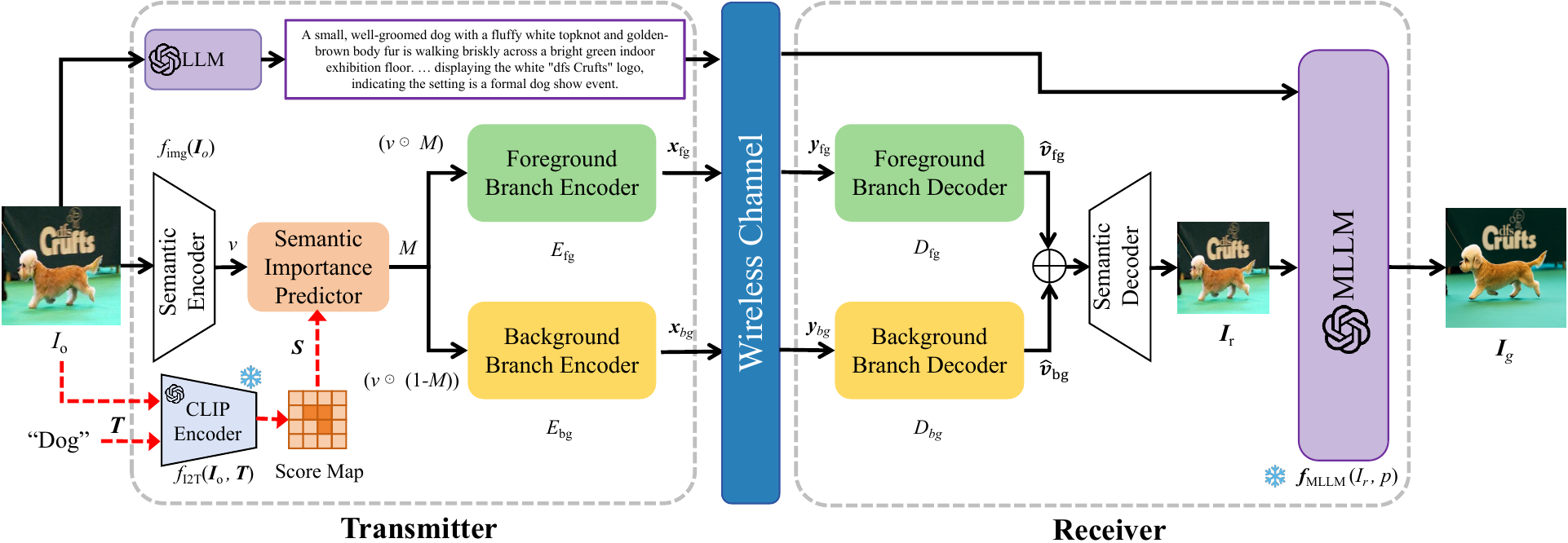}
	\caption{\small{ {Illustration of the proposed system model. Red dash line indicates training-only. }  {The captioning and generative editing blocks denote generic model functions; their concrete implementations are specified in Sections~\ref{sec:main_backend} and~\ref{sec:edge_deployment}.}}}
	\label{fig1}
\end{figure*}

\begin{figure}[!t]   
	\centering        
	\includegraphics[clip,width=0.9\linewidth]{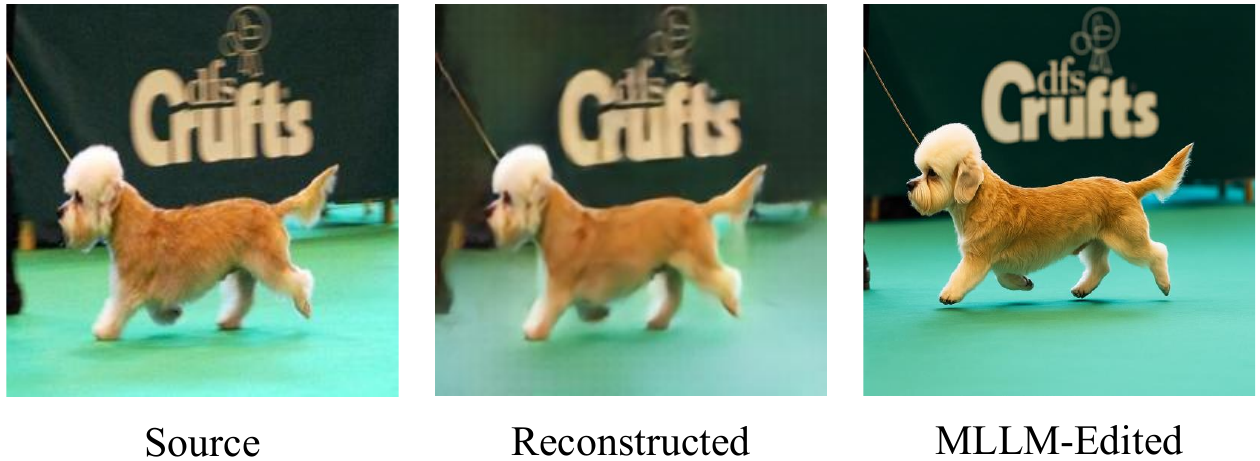}
	\caption{\small{Visualization zoom-in for Fig.~\ref{fig1}.}}
	\label{fig_sample}
\end{figure}

     {As illustrated in Fig.~\ref{fig1}, we consider a single-user semantic image transmission system. The transmitter extracts two types of semantic information from the input image $\bm{I}_\mathrm{o}\in \mathbb{R}^{C \times H \times W}$: visual features for transmission and semantic importance scores derived from a  {pretrained image--text alignment model}. These signals are jointly used to route different image regions through foreground and background branches before transmission over a noisy wireless channel. At the receiver, the decoded branch features are fused into an auxiliary image $\bm{I}_\mathrm{r}\in \mathbb{R}^{C \times H \times W}$. The auxiliary image and a transmitted prompt then condition a  generative editing backend to produce the final image $\bm{I}_\mathrm{g}\in \mathbb{R}^{C \times H \times W}$.}

\subsection{Transmitter}

     {The transmitter consists of three main components: a visual feature encoder, an image--text-alignment-based semantic routing module, and a dual-branch encoder.} 
    The objective is to retain semantically critical regions while minimizing the bandwidth cost.

\subsubsection{Visual Feature Extraction Path}
    The input R.G.B image $\bm{I}_\mathrm{o}$ is first embedded via a convolutional encoder to extract patch-level semantic features:
    \begin{equation}
        \bm{v} = f_\text{img}(\bm{I}_\mathrm{o}),
    \end{equation}
    where $f_\text{img}(\cdot)$ denotes the {semantic} encoder and $\bm{v}\in \mathbb{R}^{C' \times H' \times W'}$ is the intermediate visual representation, which serve as the input to the semantic importance predictor and as the source features for subsequent transmission.
    Notably, $C'$, $H'$ and $W'$ represent dimension, height, and width of the extracted feature map. 

\subsubsection{Semantic Routing Path}
    During training, a pretrained image--text alignment encoder $f_\mathrm{I2T}$, such as CLIP\footnote{\url{https://github.com/openai/CLIP}}~\cite{clip}, generates a teacher score map from the input image $\bm{I}_\mathrm{o}$ and a class-level textual label $T$ (e.g., ``a photo of a dog''):
\begin{equation}
	\bm{S}^{\mathrm{tea}} = f_\mathrm{I2T}(\bm{I}_\mathrm{o}, T),
	\label{eq:teacher_score}
\end{equation}
where $\bm{S}^{\mathrm{tea}} \in \mathbb{R}^{H' \times W'}$ represents spatial image--text alignment and is used only as a training-time teacher signal. The semantic importance predictor (SIP) estimates the importance mask directly from the visual feature map:
\begin{equation}
	\bm{M} = f_\mathrm{SIP}(\bm{v}),
	\label{eq:sip_abstract}
\end{equation}
where $\bm{M} \in [0,1]^{H' \times W'}$ indicates the semantic relevance of each spatial patch. At inference, neither $T$ nor $\bm{S}^{\mathrm{tea}}$ is required.

\subsubsection{Dual-Branch Routing and Encoding}
    Based on $\bm{M}$, the visual feature $\bm{v}$ is split into two spatial groups: semantic-important (i.e., semantic foreground) and non-semantic-important (i.e., semantic background) patches. 
    The two branches are encoded separately:
    \begin{align}
        \bm{x}_\text{fg} &= \mathrm{E}_{\text{fg}}(\bm{v} \odot \bm{M}), \\
        \bm{x}_\text{bg} &= \mathrm{E}_{\text{bg}}(\bm{v} \odot (1 - \bm{M})),
    \end{align}
    where $\odot$ denotes element-wise multiplication. 
   {The foreground encoder $\mathrm{E}_{\text{fg}}(\cdot)$ is implemented as a Swin Transformer module, which is chosen for its ability to capture hierarchical and long-range contextual information, thereby preserving semantic richness and structural details essential for accurate foreground representation. On the other hand, the background encoder {$\mathrm{E}_{\text{bg}}(\cdot)$} employs a convolutional architecture augmented with the Convolutional Block Attention Module (CBAM) \cite{Woo_2018_ECCV}, following the design in \cite{choi2025feature}. This choice is motivated by the need to efficiently compress redundant or less informative regions in the background, while still emphasizing repetitive patterns and broader contextual features rather than fine-grained texture details. It is worth noting that the specific architectural choice for the background encoder is not a core contribution of our work, and alternative modules with similar compression and attention capabilities could be substituted. {The Swin Transformer and CBAM are employed as effective, well-understood components that satisfy the distinct functional requirements (detailed foreground preservation vs. efficient background compression) within our GenEditSC paradigm.}}
    The outputs are transmitted to the receiver through wireless channels. 

    \textbf{Note:}  {As shown in Fig.~\ref{fig1}, an image-captioning model generates a descriptive prompt $p=f_\mathrm{cap}(\bm{I}_\mathrm{o})$, which is compressed and transmitted as an auxiliary condition to the receiver. In the main experiments, $f_\mathrm{cap}$ is implemented using GPT-4o; the model-specific settings are reported in Section~\ref{sec:main_backend}.} 
    
\subsection{Wireless Channel}

    For simplicity, we assume that the encoded foreground and background features share the same wireless channel. The transmitted signal $\bm{x}\in\mathbb{R}^{d}$, where $d=C'\times H'\times W'$, is
    \begin{equation}
        \bm{x}=\operatorname{Concat}(\bm{x}_\mathrm{fg},\bm{x}_\mathrm{bg}).
    \end{equation}
    The received signal $\bm{y}\in\mathbb{R}^{d}$ is modeled as
    \begin{equation}
        \bm{y}=\bm{h}\odot\bm{x}+\bm{n},\qquad \bm{n}\sim\mathcal{N}(\bm{0},\sigma^2\bm{I}),
    \end{equation}
    where $\odot$ denotes element-wise multiplication, $\bm{h}\in\mathbb{R}^{d}$ is the channel-gain vector, and $\sigma^2$ is the noise power. We consider AWGN and Rayleigh fading channels. For the AWGN channel, $\bm{h}=\bm{1}$. For the Rayleigh fading channel, each amplitude coefficient is modeled as $h_i=|g_i|$ with $g_i\sim\mathcal{CN}(0,1)$, so that $h_i$ follows a Rayleigh distribution; the additive-noise model remains unchanged.

Similar to the existing work of generative semantic communication,  the text prompt is conveyed through a conventional digital communication link. At a high level, the recovered prompt is represented as
\begin{equation}
	\hat{p}=f_{\mathrm{txt,dec}}\!\left(\mathcal{H}_{\mathrm{txt}}\!\left(f_{\mathrm{txt,enc}}(p)\right)\right),
	\label{eq:text_link}
\end{equation}
where $f_{\mathrm{txt,enc}}(\cdot)$, $\mathcal{H}_{\mathrm{txt}}(\cdot)$, and $f_{\mathrm{txt,dec}}(\cdot)$ denote digital encoding, the text-link channel, and digital decoding, respectively. The concrete source encoding, channel coding, and modulation settings are provided in Section~V-A4.

\subsection{Receiver}
    On the receiver side, the received signal $\bm{y}$ is decomposed to retrieve the noisy foreground features $\bm{y}_\mathrm{fg}$ and the background features $\bm{y}_\mathrm{bg}$, which are fed into two specialized decoders to reconstruct semantic features of the two branches. 
     {The foreground decoder $\mathrm{D}_\text{fg}(\cdot)$ and background decoder $\mathrm{D}_\text{bg}(\cdot)$ are respectively given by}
    \begin{align}
        \hat{\bm{v}}_\text{fg} &= \mathrm{D}_\text{fg}(\bm{y}_\text{fg}), \\
        \hat{\bm{v}}_\text{bg} &= \mathrm{D}_\text{bg}(\bm{y}_\text{bg}).
    \end{align}
   {The coarse reconstructed image is given by }
    \begin{equation}
        \bm{I}_\mathrm{r} = f_\text{fusion}(\hat{\bm{v}}_\text{fg}, \hat{\bm{v}}_\text{bg}).
    \end{equation}
     {Finally, the auxiliary image $\bm{I}_\mathrm{r}$ and the prompt $p$ are fed into a generative editing backend implemented by an MLLM:}
    \begin{equation}
        \bm{I}_\mathrm{g} = f_\text{MLLM}(\bm{I}_\mathrm{r}, \hat{p}),
    \end{equation}
     {which produces the final edited image $\bm{I}_\mathrm{g}$. Here, $f_\text{MLLM}$ denotes a  {fixed} and replaceable generative editing function.  {In this paper, ``fixed'' means that the backend is not updated during optimization of the JSCC front-end; the backend itself may be instantiated by a pretrained model.} The proposed JSCC architecture and training objective do not require joint optimization with a particular backend implementation.}

\section{Network Architecture}\label{basic_model_design}
\subsection{CLIP-Based Cross-Modal Context Matching Module}
	\begin{figure}[!t]   
		\centering        
		\includegraphics[clip,width=\linewidth]{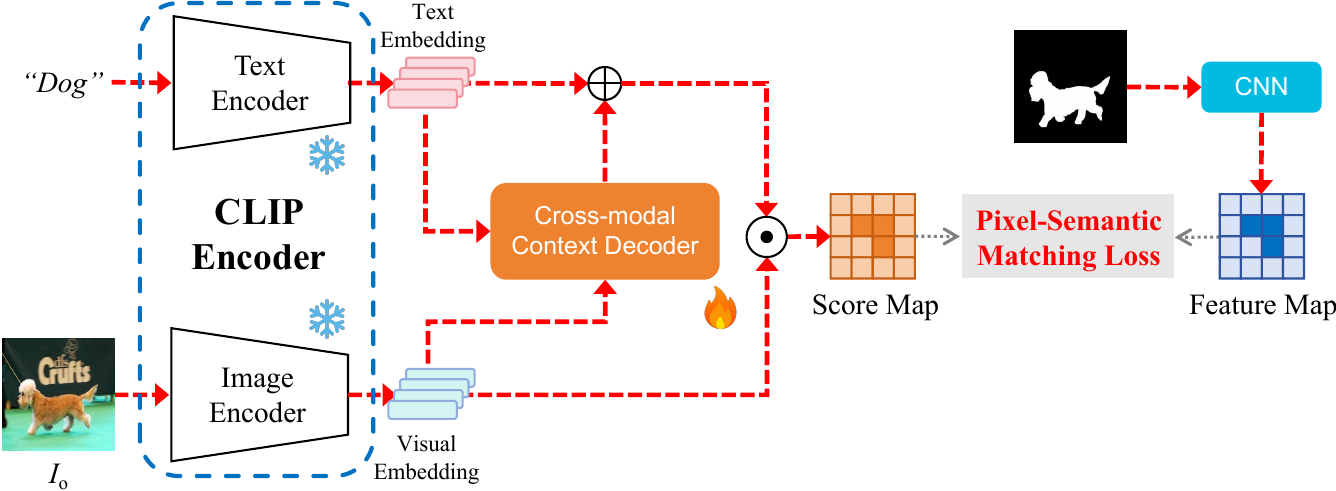}
		\caption{\small{ {Score map generation via CLIP-based cross-modal matching.}}}
		\label{fig2}
	\end{figure}

	As shown in Fig.~\ref{fig2}, given an input image $\bm{I}_\mathrm{o}$ and a target class text $T$ (e.g., ``Dog''), we utilize a  {pretrained} CLIP model~\cite{denseclip} to extract semantic-aligned visual and textual features. 
	Specifically, the CLIP image encoder outputs the global feature $\bm{g} \in \mathbb{R}^{C}$ and the patch-level visual embedding $\bm{e}_v \in \mathbb{R}^{C \times H \times W}$. The CLIP text encoder generates the class-level text embedding $\bm{e}_\mathrm{t} \in \mathbb{R}^{C}$.

    To bridge the gap between generic text descriptions and pixel-level image content, we refine $\bm{e}_\mathrm{t}$ into a context-aware representation $\tilde{\bm{e}}_\mathrm{t}$ by fusing it with the aggregated visual context $[\bm{g}; \bm{e}_\mathrm{v}]$ via a cross-modal transformer decoder. This process yields a refined semantic vector that is specifically attuned to the input image, effectively pinpointing the latent visual attributes that correspond to the text prompt $T$.

    The similarity between $\tilde{\bm{e}}_\mathrm{t}$ and each spatial location in $\bm{e}_v$ is then calculated to generate the semantic importance score map $\bm{S} \in \mathbb{R}^{H \times W}$. This yields a spatially varying prior that highlights semantically salient regions corresponding to $T$, providing precise, pixel-level guidance for the subsequent routing module. Complete computational details are delineated in Appendix A.

\subsection{Semantic Importance-wise Dual-Branch Feature Encoding}
\subsubsection{Semantic Importance Prediction Module}
	To enable effective semantic-aware encoding under channel constraints, we introduce a \textit{Semantic Importance Prediction Module} (SIPM) that adaptively identifies and enhances foreground patches while reducing redundancy in background regions. Specifically, given the encoder feature map $\bm{v} \in \mathbb{R}^{C \times H \times W}$, a semantic score map $\bm{S} \in \mathbb{R}^{1 \times H \times W}$ derived from CLIP, and a global modulation map $\bm{A}_\mathrm{g} \in \mathbb{R}^{2 \times H \times W}$ obtained by passing $\bm{v}$ through a shallow CNN, we aggregate them to construct the routing condition tensor:
	\begin{equation}
		\bm{e}_{\text{cond}} = \operatorname{Concat}(\bm{v}, \bm{A}_\mathrm{g}, \hat{S}) \in \mathbb{R}^{(C + 3) \times H \times W},
	\end{equation}
	where $\hat{S}$ denote the prediction output of $f_{SIP}$. 
	Notably, $\bm{e}_\mathrm{cond}$ integrates structural features, semantic saliency, and global modulation cues. 
	Moreover, $\bm{e}_\mathrm{cond}$ is fed into the semantic predictor network $\Psi$ to facilitate reliable patch-wise discrimination. 
	The network produces four outputs{, as shown in Fig.~4}:
	\begin{equation}
		\bm{M}, \Delta p, \bm{A}_{\text{CA}}, \bm{A}_{\text{SA}} = \Psi(\bm{e}_{\text{cond}}),
	\end{equation}
	where $\bm{M} \in \{0,1\}^{N \times 1}$ is a binary semantic mask that classifies each patch as foreground ($\bm{M}_{i,j}=1$) or background ($\bm{M}_{i,j}=0$). 
	The offset field $\Delta p \in \mathbb{R}^{2 \times H \times W}$ predicts dense deformable displacements for sampling alignment. $\bm{A}_{\text{CA}} \in \mathbb{R}^{C \times 1 \times 1}$ is a global channel attention map, and $\bm{A}_{\text{SA}} \in \mathbb{R}^{1 \times H \times W}$ is a spatial attention map that modulates background residuals.

	Note that in contrast to the other outputs, which operate at pixel or channel-level granularity, $\bm{M}$ is computed at the patch level to facilitate lightweight semantic separation. Specifically, to compute $\bm{M}$, the condition tensor $\bm{e}_{\text{cond}}$ is partitioned into $N = \frac{H \cdot W}{w^2}$ non-overlapping spatial patches of size $w \times w$. 
	Each patch is average pooled to obtain a compact patch descriptor $\bm{e}_{\text{patch}} \in \mathbb{R}^{N \times w^2}$. 
	A shallow MLP $\mathcal{F}(\cdot)$ followed by Gumbel Softmax $\mathcal{G}(\cdot)$ yields the binary mask:
	\begin{equation}
		\tilde{\bm{M}} =  \operatorname{GumbelSoftmax}(\mathcal{F}(\bm{e}_{\text{patch}})).
	\end{equation}
The relaxed mask $\widetilde{\bm{M}}$ provides a differentiable implementation of the abstract mask $\bm{M}$ defined in Section~III. At inference, the hard foreground/background assignment $\bm{M}$ is obtained from the maximum-logit class for each patch.

	\begin{figure}[!t]
	\centering
	\begin{minipage}{0.8\linewidth}
		\centering
		\includegraphics[width=\linewidth]{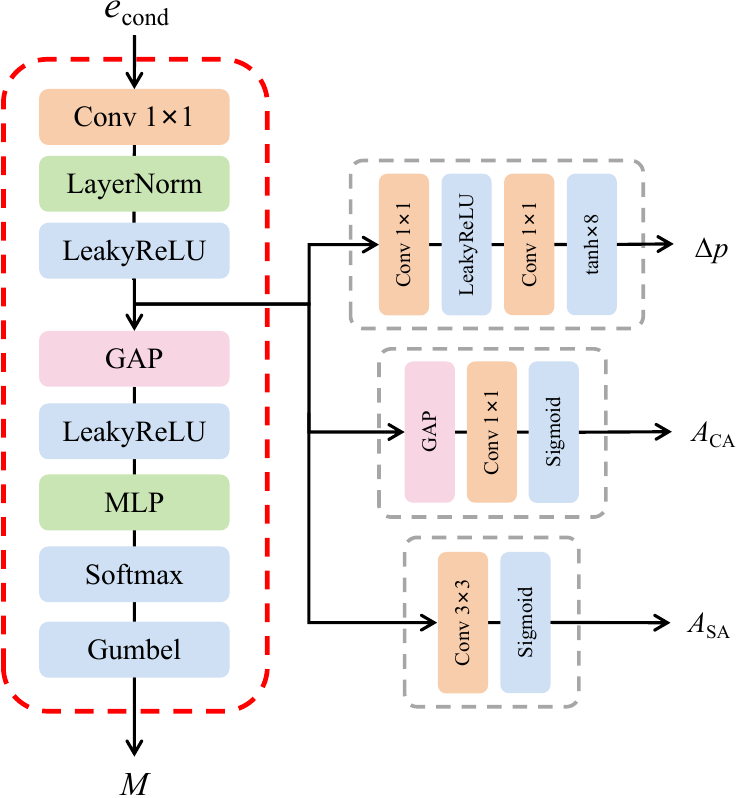}
		\caption{\small{Semantic Importance Prediction Module.}}
		\label{fig3}
	\end{minipage}
	\end{figure}

\subsubsection{Foreground-Background Split Attention}
	\begin{figure*}[!t]   
		\centering        
		\includegraphics[clip,width=0.99\linewidth]{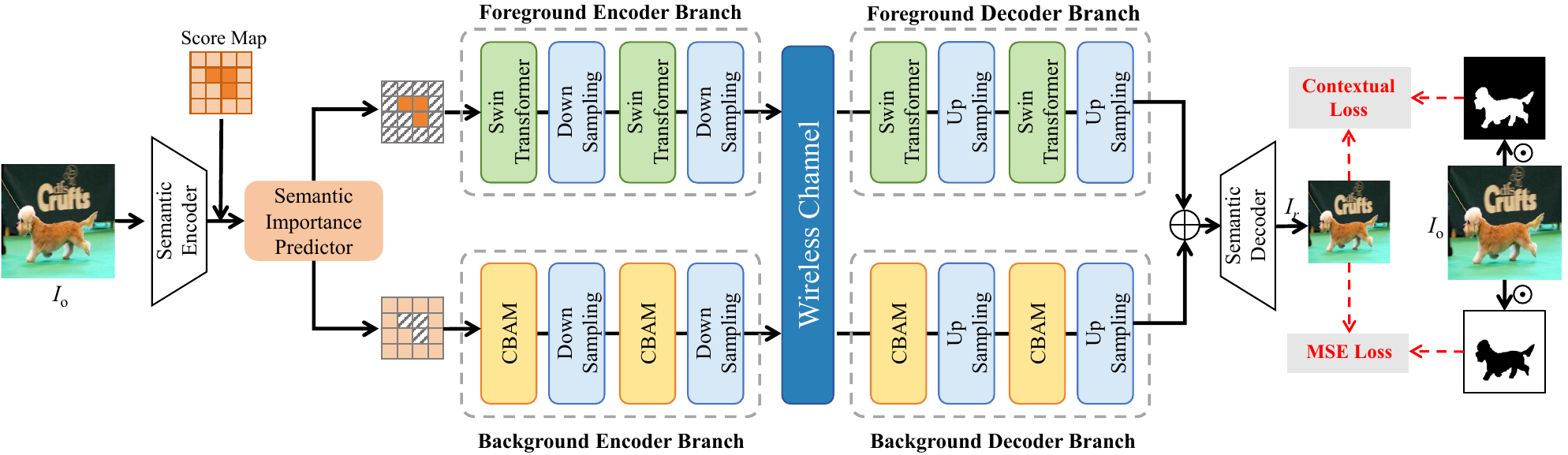}
		\caption{\small{Illustration of the Semantic Importance-wise Dual-Branch. }}
		\label{fig4}
	\end{figure*}

	The foreground branch uses Swin Transformer blocks to preserve long-range dependencies and structural details in semantically important regions. The background branch uses a lightweight ResConv-CBAM architecture to retain coarse contextual information with lower representational complexity. Their branch representations are written as
	\begin{align}
		\bm{F}_{\text{fg}} &= \phi_{\text{Swin}}(\bm{v} \odot \bm{M}), \\
		\bm{F}_{\text{bg}} &= \phi_{\text{CBAM}}(\bm{v} \odot (1-\bm{M})).
	\end{align}
	The two branch features are restored to a common spatial grid and fused by element-wise addition:
	\begin{equation}
		\bm{F}_{\text{out}} = \bm{F}_{\text{fg}} + \bm{F}_{\text{bg}}.
	\end{equation}
	As illustrated in Fig.~\ref{fig4}, the corresponding decoder branches use the same asymmetric functional roles: the Swin-based branch preserves foreground structure, whereas the ResConv-CBAM branch reconstructs background context.


{The asymmetric dual-branch design applies different feature-processing roles to foreground and background content within the limited transmission representation.}

\subsection{Training and Loss Function Design}

	The fixed generative editing backend is excluded from front-end training. Therefore, the proposed loss does not directly optimize the final edited image $\bm{I}_\mathrm{g}$. Instead, it trains the JSCC front-end to produce an intermediate reconstruction $\bm{I}_\mathrm{r}$ that preserves foreground details and background context, thereby providing a more effective visual condition for downstream editing.

\subsubsection{Pixel-Semantic Matching Loss}
{To improve localization of semantically important regions and supervise the training-time teacher score map $\bm{S}^{\mathrm{tea}}$, we introduce a \emph{pixel--semantic matching loss}.}  
Since no ground-truth semantic importance annotations are available, we adopt precomputed saliency maps $\bm{M}_{\text{sal}}\in \{0,1\}^{H \times W}$ as pseudo-labels to supervise the model in a weakly supervised manner.
{To focus the supervision on salient regions, the loss is evaluated only at foreground positions $(\bm{M}_{\text{sal}}=1)$:}
\begin{equation}
	\mathcal{L}_{\text{match}} =
	\frac{
		\sum_{i,j} \Big[ \bm{M}_{\text{sal}}(i,j) \cdot \operatorname{BCE}\big(\sigma(\bm{S}^{\mathrm{tea}}(i,j)), \bm{M}_{\text{sal}}(i,j)\big) \Big]
	}{
		\sum_{i,j} \bm{M}_{\text{sal}}(i,j) + \epsilon
	},
\end{equation}
{where $\sigma(\cdot)$ denotes the sigmoid activation, $\operatorname{BCE}(\cdot)$ is the element-wise binary cross-entropy loss, and $\epsilon$ is a small constant for numerical stability. This foreground-masked BCE encourages high teacher-map responses on salient regions while excluding uncertain background positions from the matching loss.}  

\subsubsection{Foreground Reconstruction Loss}
To preserve fine-grained semantic information in the reconstructed image, we impose a reconstruction constraint on the salient regions indicated by the saliency pseudo-label $\bm{M}_{\text{sal}} \in \{0,1\}^{H \times W}$. 
We use a pixel-wise mean squared error (MSE) loss, computed only over foreground pixels where $\bm{M}_{\text{sal}}(i,j) = 1$:
\begin{equation}
	\mathcal{L}_{\text{fg}} =
	\frac{
		\sum_{i,j} \bm{M}_{\text{sal}}(i,j) \cdot \left\| \bm{I}_r(i,j) - \bm{I}_o(i,j) \right\|^2
	}{
		\sum_{i,j} \bm{M}_{\text{sal}}(i,j) + \epsilon
	},
\end{equation}
where $\bm{I}_o, \bm{I}_r \in \mathbb{R}^{3 \times H \times W}$ denote the original RGB images and the intermediate reconstructed {images} from {GenEditSC}, and $\epsilon$ is a small constant to avoid division by zero. 
{This foreground-weighted MSE assigns greater reconstruction emphasis to semantically important regions, such as salient objects and foreground entities.}

\subsubsection{Background Contextual Loss}

{While foreground regions require detailed reconstruction, background areas often contain repetitive or less informative content. We therefore adopt a \textit{contextual loss} that preserves their overall appearance and structure while allowing fine-grained texture simplification.}

Given the reconstructed image $\bm{I}_r$ and the original image $\bm{I}_o$, we first extract deep features using a  {pretrained} VGG network: 
\begin{equation}
	f_r = \Phi(\bm{I}_r), \quad f_o = \Phi(\bm{I}_o),
\end{equation}
where $f_r, f_o \in \mathbb{R}^{C' \times H' \times W'}$ are the extracted feature maps. The saliency pseudo-label map $\bm{M}_{\text{sal}}$ is downsampled to match the feature resolution, yielding $\bm{M}'_{\text{sal}} \in \{0,1\}^{H' \times W'}$. 
We then compute the contextual loss only over background positions where $\bm{M}'_{\text{sal}}(i,j) = 0$:
\begin{equation}
	\mathcal{L}_{\text{bg}} =
	\frac{
		\sum_{i,j} \left[ (1 - \bm{M}'_{\text{sal}}(i,j)) \cdot \mathcal{CX}\left(f_r(i,j), f_o(i,j)\right) \right]
	}{
		\sum_{i,j} (1 - \bm{M}'_{\text{sal}}(i,j)) + \epsilon
	},
\end{equation}
where $\mathcal{CX}(\cdot, \cdot)$ denotes the contextual similarity:
\begin{equation}
	\mathcal{CX}(x, y) = -\log \left(
	\frac{
		\exp\left( \operatorname{sim}(x, y)/\tau \right)
	}{
		\sum_{y'} \exp\left( \operatorname{sim}(x, y')/\tau \right)
	}
	\right),
\end{equation}
where $\operatorname{sim}(\cdot, \cdot)$ is the cosine similarity between feature vectors, and $\tau$ is a temperature hyperparameter. 
This background-weighted contextual loss enables flexible degradation of textures while maintaining perceptual consistency in non-salient regions.

\subsubsection{Overall Loss}
The final training objective is a weighted sum of the above three losses:
\begin{equation}
	\mathcal{L}_{\text{total}} = \lambda_1 { \mathcal{L}_{\text{match}}} + \lambda_2 \mathcal{L}_{\text{fg}} + \lambda_3 \mathcal{L}_{\text{bg}},
\end{equation}
where $\lambda_1$, $\lambda_2$, and $\lambda_3$ are hyperparameters that balance the contributions of semantic alignment, target fidelity, and efficient background compression.

\textbf{Note:} {The saliency maps $M_{\text{sal}}$, the class label $T$, and the teacher score map $\bm{S}^{\mathrm{tea}}$ are used only during training. Inference requires neither external annotations nor saliency inputs.}
\subsection{ {Generative Editing for Image Reconstruction}}
	 {To exploit pretrained multimodal generative models for semantic enhancement, we adopt a generative editing pipeline that operates on the JSCC-reconstructed image $\bm{I}_\mathrm{r}$. Given the imperfect but semantically aligned reconstruction produced by the GenEditSC decoder, the editing backend is instructed to preserve the target foreground and scene layout while refining missing or degraded details according to a text prompt.}
	 {We first use the following instruction to generate a detailed image caption at the transmitter:}
	\begin{center}
	\fbox{%
	\parbox{0.9\linewidth}{%
	\small
	\textbf{Prompt:} \\
	\emph{Describe the image in precise detail to enable accurate text-to-image generation. First, summarize the overall scene and tone. Next, identify all main objects, including their quantities, colors, and appearances. Then, specify the absolute position (e.g., top, bottom, left, right) of each object and the relative spatial relationships between them (e.g., above, below, in front of, behind). Finally, describe how these objects are positioned within the background scene. Provide a comprehensive, detailed, and accurate description to ensure the generated image faithfully reconstructs the original.}
	}
	}
	\end{center}
	
	 {We then use a task-specific prompt to guide the generative editing backend at the receiver:}
	\begin{center}
	\fbox{%
	\parbox{0.9\linewidth}{%
	\small
	\textbf{Prompt:} \\
	\emph{This is a compressed image. Please enhance the resolution and sharpness of this image while preserving its original colors, details, composition, and layout. The restored image should look clear and high-definition. Please control the brightness and average brightness of each part of the generated image to be consistent with the reference image I gave. Use the following text to reconstruct: \{Caption from the image\}.}
	}
	}
	\end{center}

	 {This prompt instructs the model to preserve foreground identity, colors, composition, and layout while refining the remaining content from the textual description. Because no explicit editing mask is supplied to the backend in Eq.~(11), foreground preservation is a behavioral constraint induced by the reconstructed image and the prompt rather than a hard mask-constrained operation.}
	 {This design reduces information redundancy in transmission by selectively encoding semantically important content through JSCC, while reconstructing less detailed regions through the text-guided generative editing backend.}

	 {By combining semantic-aware compression with text-conditioned editing, the framework preserves critical image regions while allowing flexible refinement of missing details, supporting bandwidth-efficient semantic image communication.}

\section{ {Experimental Results}}\label{simulation_results}
     {In this section, we evaluate the proposed method by addressing the following questions:}
    \begin{itemize}
        \item  {Compared with discriminative JSCC, does GenEditSC improve semantic preservation and perceptual quality, particularly under low-SNR channel conditions? (Section~\ref{sec:disc_results})}
        \item  {Compared with generative SemComm, does GenEditSC improve semantic preservation, perceptual quality, and visual fidelity under matched experimental settings? (Section~\ref{sec:gen_results})}
        \item  {Does GenEditSC achieve a favorable trade-off between semantic reconstruction quality and communication overhead? (Section~\ref{sec:overhead_results})}
        \item  {Can the generative editing backend be executed on an embedded edge platform with measurable memory, latency, and reconstruction quality? (Section~V.D)}
        \item  {How much does each proposed component and the asymmetric dual-branch architecture contribute to the overall performance? (Section~\ref{sec:ablation})}
    \end{itemize}
     {The qualitative results in Section~\ref{sec:qualitative_results} further illustrate the structural and perceptual differences observed in the first two comparisons.}
    
\subsection{ {Experimental Setup}} 
\subsubsection{Datasets}
We use the DUTS salient object detection dataset and the Pascal VOC dataset as the experimental data for training and testing. The DUTS dataset includes DUTS-TR with 10,553 images for training and DUTS-TE with 5,019 images for testing. The Pascal VOC dataset consists of 10,000 images, divided into training and testing sets with a 9:1 ratio.
All images are resized to $256\times256$ pixels during pre-processing. 
The ground-truth saliency masks are manually annotated to highlight the most visually prominent objects in each scene, guided by human visual attention. 

 {Notably, DUTS contains diverse scenes, including cases with multiple prominent objects, rather than only single-object images. Therefore, the learned semantic importance predictor is trained with saliency patterns beyond simple single-object settings.} The DUTS-TR annotations serve solely as weak semantic priors for training our semantic importance prediction module.  {During inference, the semantic importance predictor directly estimates the foreground/background mask from the input features, without requiring saliency annotations or the class-level text label $T$.}

\subsubsection{ {JSCC Training Settings}}
     {We use the AWGN channel as the default training channel and additionally evaluate Rayleigh fading where specified. During training, the channel SNR in each iteration is sampled from \{1, 4, 7, 10\}~dB.}
      {Separate models are trained for $\mathrm{CR}\in\{1/48,1/36,1/24,1/12\}$, with $\mathrm{CR}=1/24$ used as the default operating point to align with SGDJSCC~\cite{sgdjscc}.}
      {The trainable communication modules---including the visual feature encoder, cross-modal context decoder, semantic importance predictor, dual-branch encoders, and corresponding decoders---are optimized jointly without separate foreground/background training stages.}
      {The pretrained CLIP image and text encoders and the pretrained VGG feature extractor are kept frozen throughout training, and the fixed generative editing backend is excluded from optimization.}
     {The model is trained for 2000 epochs using Adam with a learning rate of $10^{-4}$ on an NVIDIA RTX~4090 under Ubuntu~20.04 LTS with an Intel\textregistered\ Xeon\textregistered\ Platinum~8336C CPU at 2.30~GHz.}

     {The composite-loss weights are $\lambda_1=1.0$, $\lambda_2=0.0001$, and $\lambda_3=0.01$. The Gumbel--Softmax temperature is 1.0, and the contextual-loss temperature is $\tau=0.5$. Unless otherwise stated, these values define the default training configuration.}

\subsubsection{ {Captioning and Generative Editing Backend Settings}}\label{sec:main_backend}
     {In the main evaluation, GPT-4o (OpenAI, snapshot \texttt{gpt-4o-2024-05-13}) is used to generate the descriptive prompt $p$ and serves as the primary experimental implementation of the receiver-side generative editing function $f_\mathrm{MLLM}$.}
      {GPT-4o is used as a fixed, inference-only backend and is not jointly optimized with the JSCC front-end.}
     {The captioning and editing instructions are given in Section~IV-D.}

   \begin{figure}[!t]  
	\centering
	\includegraphics[clip,width=0.85\linewidth]{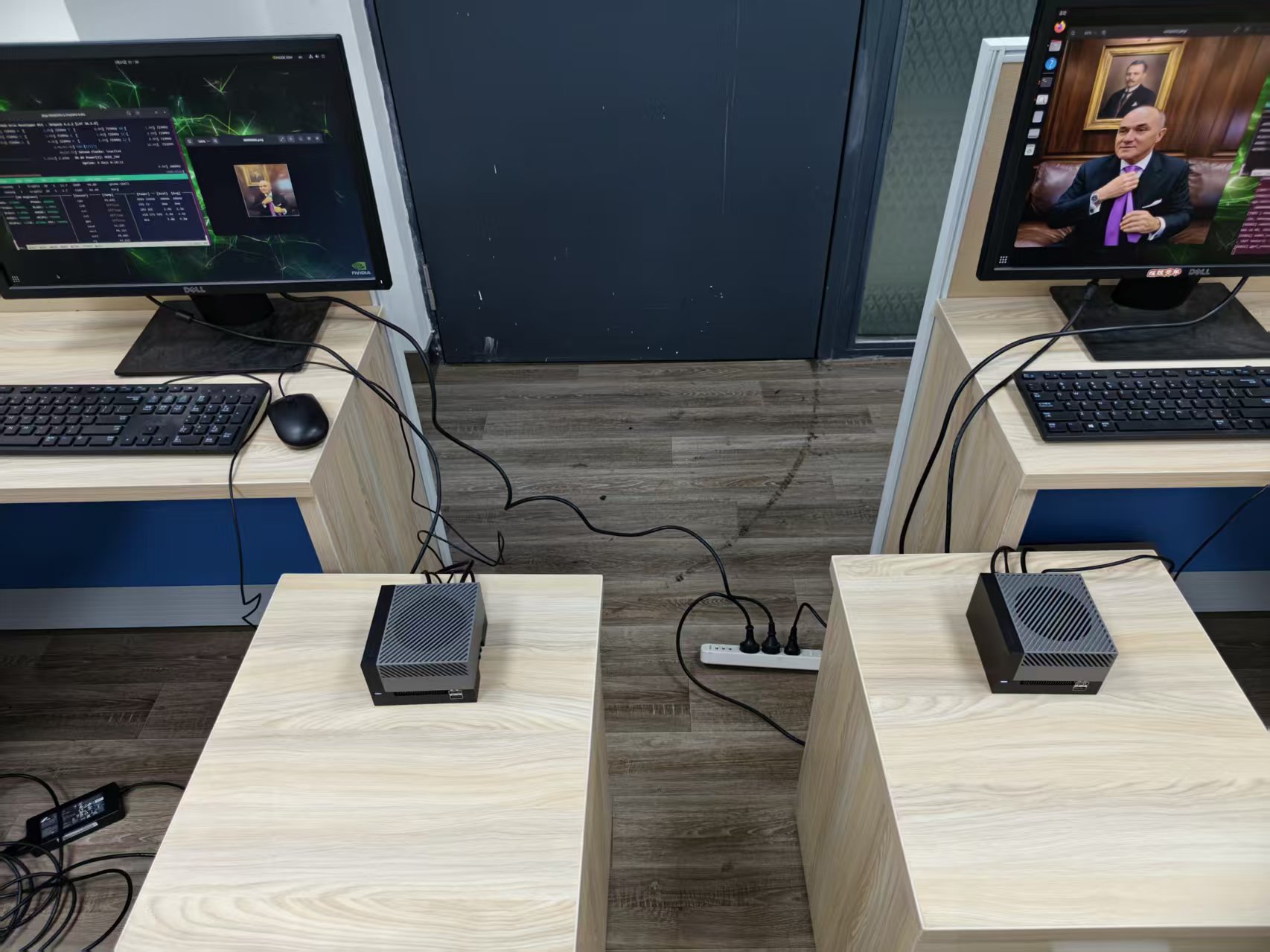}
	\caption{ {\small{Local Hardware Testbed.}}}	\label{tab:testbed}
\end{figure}
\begin{figure*}[!t]
	\centering
	\begin{subfigure}[b]{0.325\textwidth}
		\includegraphics[width=\linewidth]{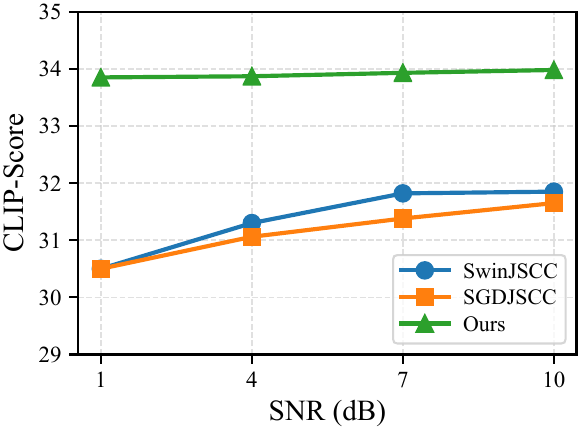}
		\subcaption{CLIP-Score on (DUTS, AWGN).}
		\label{fig:duts_clip_awgn}
	\end{subfigure}
	\begin{subfigure}[b]{0.325\textwidth}
		\includegraphics[width=\linewidth]{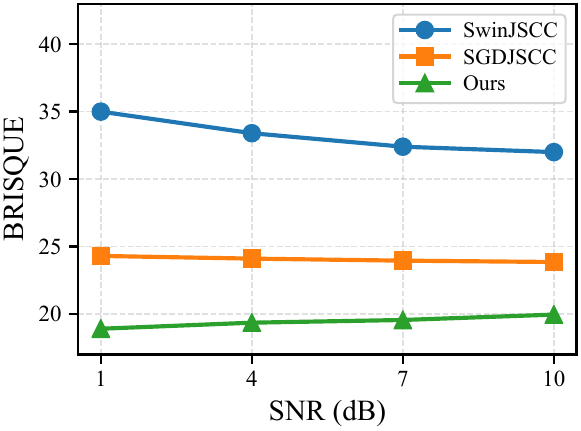}
		\subcaption{BRISQUE on (DUTS, AWGN).}
		\label{fig:duts_brisque_awgn}
	\end{subfigure}
	\begin{subfigure}[b]{0.325\textwidth}
		\includegraphics[width=\linewidth]{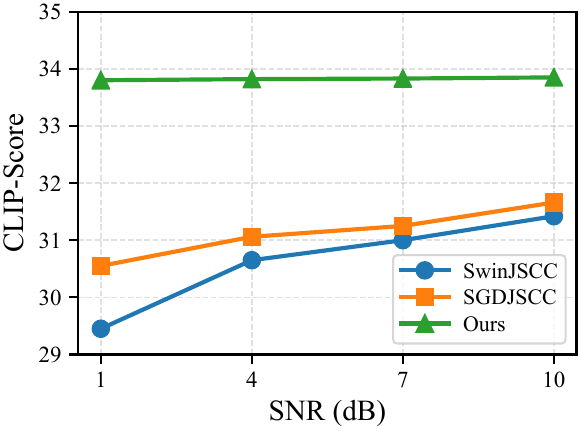}
		\subcaption{CLIP-Score on (DUTS, Rayleigh).}
		\label{fig:duts_clip_rayleigh}
	\end{subfigure}
	
	\begin{subfigure}[b]{0.325\textwidth}
		\includegraphics[width=\linewidth]{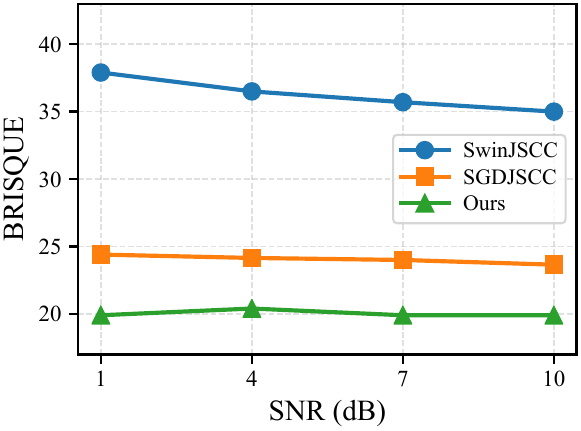}
		\subcaption{BRISQUE on (DUTS, Rayleigh).}
		\label{fig:duts_brisque_rayleigh}
	\end{subfigure}
	\begin{subfigure}[b]{0.325\textwidth}
		\includegraphics[width=\linewidth]{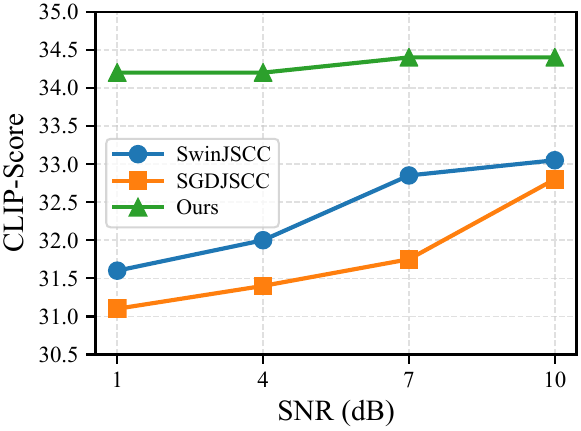}
		\subcaption{CLIP-Score on (Pascal, AWGN).}
		\label{fig:pascal_clip_awgn}
	\end{subfigure}
	\begin{subfigure}[b]{0.325\textwidth}
		\includegraphics[width=\linewidth]{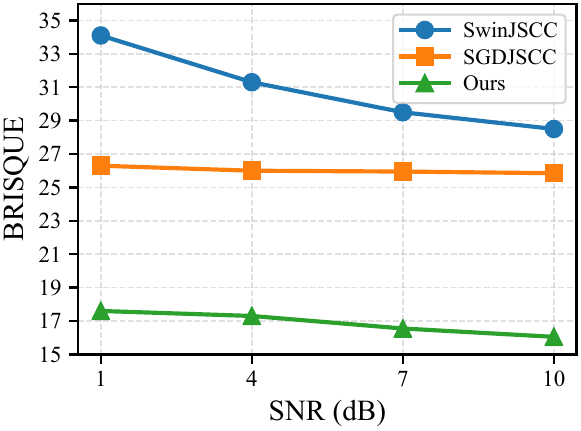}
		\subcaption{BRISQUE on (Pascal, AWGN).}
		\label{fig:pascal_brisque_awgn}
	\end{subfigure}
	
	\caption{\small Quantitative comparison with discriminative SemComm methods.}
	\label{fig:duts}
\end{figure*}
      {For the edge-deployment evaluation in Section~V.D, only the receiver-side backend $f_\mathrm{MLLM}$ is changed.}
     We deploy a 2-bit-quantized Qwen-Image model on an NVIDIA Jetson AGX Orin 64~GB. The trained JSCC model, reconstructed input $\bm{I}_\mathrm{r}$, prompt $p$, image resolution, channel realization, and evaluation samples are kept unchanged. This controlled replacement isolates the deployment characteristics of the editing backend rather than introducing a different communication scheme.

\subsubsection{ {Text Transmission Settings}}
     {To ensure a matched comparison across methods that use textual conditions, all prompts are encoded using UTF-8, converted to binary bitstreams, protected by a systematic LDPC code with rate 0.5, and mapped using 16-QAM.   {The LDPC code improves text-link reliability but does not guarantee error-free recovery. At low SNR, residual decoding errors may remain in the recovered bitstream and corrupt the UTF-8 prompt; these errors are retained in the evaluation rather than assuming successful text decoding. All prompt-based methods use the same text coding and modulation configuration to ensure a matched comparison.}}

\subsubsection{ {Communication-Overhead Accounting}}\label{sec:overhead_accounting}
      {For the comparison in Section~\ref{sec:overhead_results}, communication overhead is measured by the total number of transmitted symbols.}
      {For GenEditSC, this total includes both the analog JSCC visual stream and the LDPC-coded digital text prompt stream.}
      {Under the implementation-level accounting used in Fig.~\ref{bpp}, each transmitted element of the analog JSCC output and each 16-QAM modulation symbol is counted as one transmitted symbol.}
      {For a $3\times256\times256$ image at $\mathrm{CR}=1/24$, the visual branch requires 8192 transmitted symbols, while the text branch contributes about 426 additional modulation symbols under 16-QAM on average~\cite{11149073}, resulting in a total of 8618 symbols.}
      {For $\mathrm{CR}\in\{1/48,1/36,1/24,1/12\}$, the total transmission budgets are approximately $\{4.5\times10^3,\,5.9\times10^3,\,8.6\times10^3,\,1.68\times10^4\}$ symbols, respectively, and the corresponding text-stream shares are about $9.5\%$, $7.2\%$, $4.9\%$, and $2.5\%$.}

    

    \begin{figure*}[!t]
	\centering
	\begin{subfigure}[b]{0.32\textwidth}
		\centering
		\includegraphics[clip,width=\linewidth]{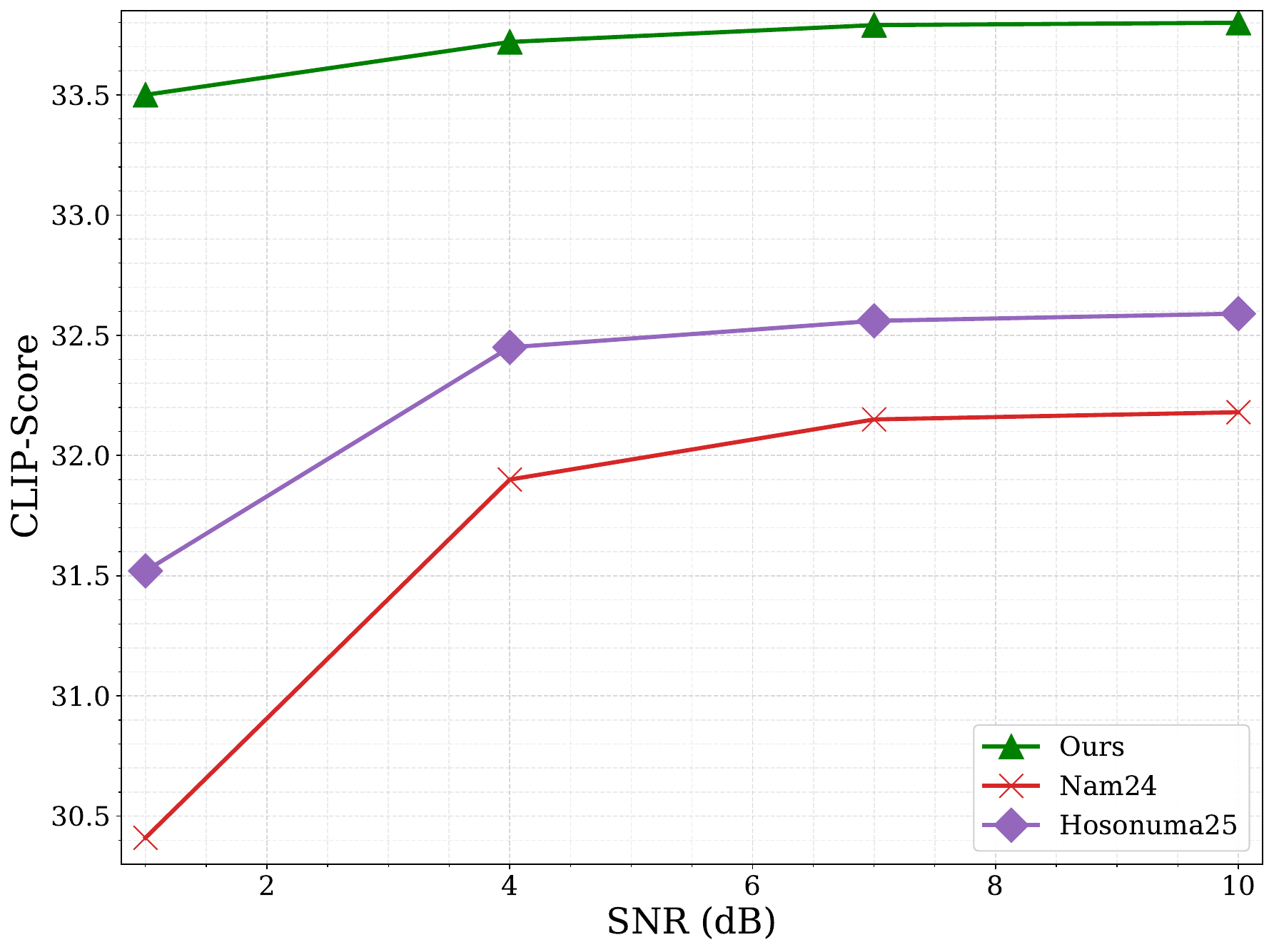}
		\caption{CLIP-Score vs SNR}
		\label{fig:clip_score}
	\end{subfigure}
	~
	\begin{subfigure}[b]{0.32\textwidth}
		\centering
		\includegraphics[clip,width=\linewidth]{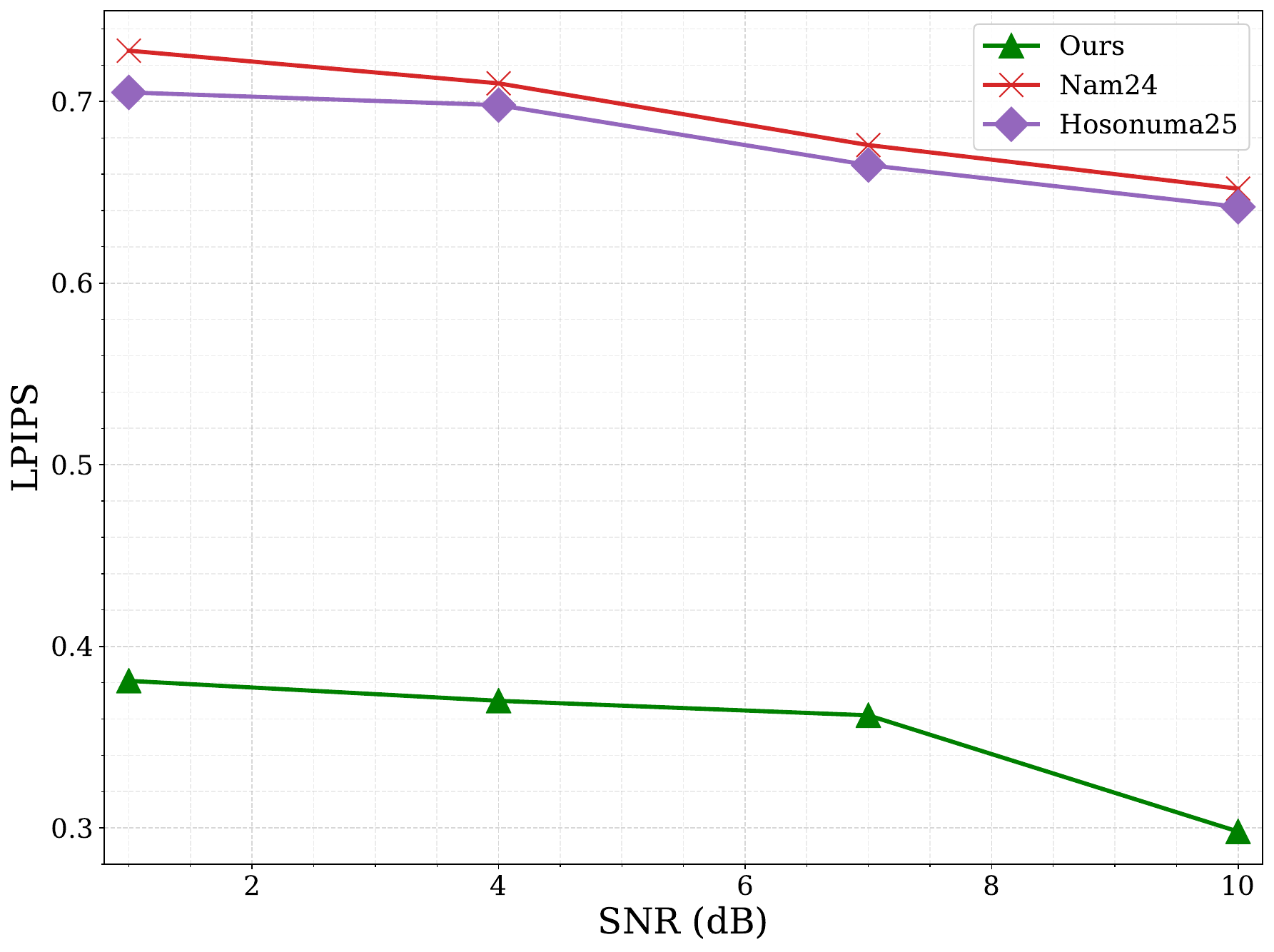}
		\caption{LPIPS vs SNR}
		\label{fig:lpips}
	\end{subfigure}
	~
	\begin{subfigure}[b]{0.32\textwidth}
		\centering
		\includegraphics[clip,width=\linewidth]{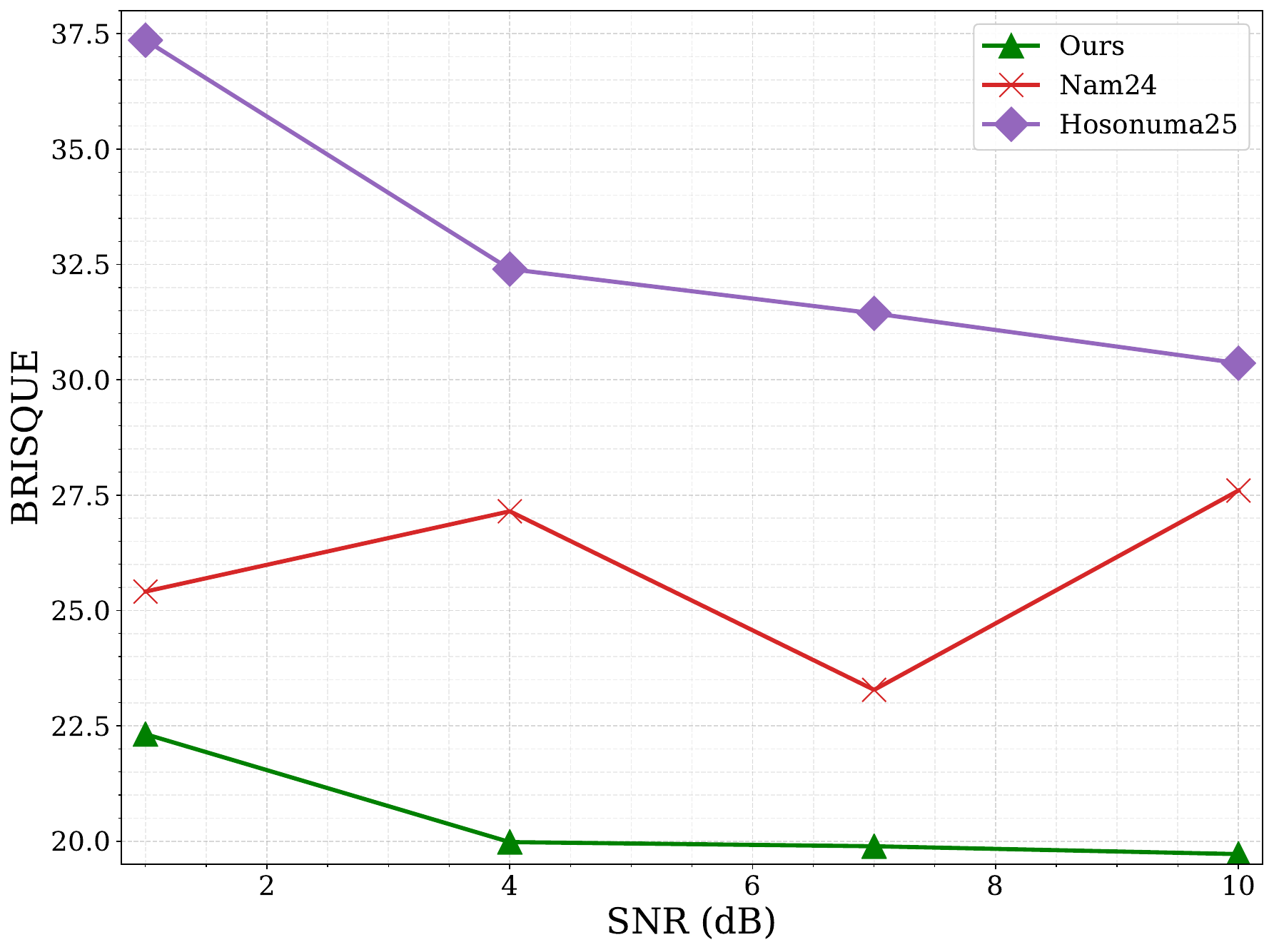}
		\caption{BRISQUE vs SNR}
		\label{fig:brisque}
	\end{subfigure}
	\caption{\small{ {Performance Comparison of Generative SemComm on DUTS dataset.}}}
	\label{fig:snr_performance}
\end{figure*}
\subsection{Benchmarks} 
    Without loss of generality, we consider the representative benchmarks,  {i.e., both discriminative and generative families,}  as follows:
    \begin{itemize}
        \item \textbf{SwinJSCC}~\cite{swinjscc}: A discriminative JSCC model that reconstructs images by transmitting dense visual features with Swin-Transformer. 
        We implement and train this baseline using the official source code\footnote{https://github.com/semcomm/SwinJSCC} on DUTS dataset, with the same training settings for fair comparison.

        \item \textbf{SGDJSCC}~\cite{sgdjscc}: A discriminative JSCC framework incorporates semantic guidance through global textual embeddings and an edge map. 
        Since the authors have not released training code but provide official inference implementation, we adopted the pre-trained model provided by the authors in our simulations.

        \item \textbf{Nam24}~\cite{jihongpark}: A generative approach that transmits only textual descriptions for reconstruction.  {We use the same caption-generation pipeline and the same main generative backend adopted in the principal GenEditSC evaluation. This control isolates differences in the transmitted semantic conditions rather than differences in backend capability.}

        \item \textbf{Hosonuma25}~\cite{mask}:  {A generative semantic communication method that additionally transmits a mask as a visual prior for text-guided image generation. We use the same caption generator and the same main generative backend as in Nam24 and GenEditSC, while accounting for the mask-transmission overhead required by this method.}
    \end{itemize}

    In the following simulations, we 
 {adopt \textbf{CLIP-Score}~\cite{clip}, \textbf{BRISQUE}~\cite{6272356}, and \textbf{LPIPS} as the evaluation metrics, which respectively measure semantic consistency, no-reference perceptual quality, and perceptual similarity. These metrics are well aligned with the objective of \emph{generative semantic communication}, where the receiver aims to reconstruct images that preserve the transmitted semantics while maintaining high perceptual quality, rather than reproducing the source image in a strictly pixel-wise manner. This choice is also consistent with recent literature. For example, \cite{MoS} employ a CLIP-based semantic metric, and \cite{canny_latency} also adopt CLIP-based evaluation for semantic reconstruction quality. Perceptual metrics are likewise widely used in related works: \cite{jihongpark} report LPIPS, while \cite{11432458} evaluate reconstruction quality using LPIPS together with other perceptual and structural measures. By contrast, conventional distortion-based metrics such as PSNR and SSIM are often less informative in generative settings, since they may penalize semantically correct and perceptually realistic outputs due to low-level pixel deviations. Therefore, the combination of CLIP-Score, BRISQUE, and LPIPS provides a more comprehensive and task-relevant evaluation of both the proposed method and the benchmark schemes.}  {Because the three metrics emphasize different properties, all performance claims are based on their joint interpretation rather than on a single score.}

\subsection{ {Quantitative Results}}   
   \subsubsection{ {Comparison with Discriminative SemComm}}\label{sec:disc_results}
     {Fig.~\ref{fig:duts} reports CLIP-Score and BRISQUE under the evaluated AWGN and Rayleigh fading conditions. GenEditSC obtains the highest CLIP-Score and the lowest BRISQUE at each tested SNR. Its metric variation over the evaluated SNR range is also smaller than that of the two discriminative baselines.}

     {As shown in Fig.~\ref{fig:duts_clip_awgn}, the average CLIP-Score of GenEditSC is $8.13\%$ and $8.89\%$ higher than that of SwinJSCC and SGDJSCC, respectively, over the AWGN channel. Under Rayleigh fading, the corresponding gains are $10.50\%$ and $8.65\%$. These results are consistent with the generation-as-editing design: the transmitted visual stream supplies a spatially anchored structural prior, while the descriptive prompt supplies complementary semantic context. Under the tested channel conditions, this combination is less sensitive to perturbations than reconstruction based only on noisy visual latent features.}

     In Fig.~\ref{fig:duts_brisque_awgn}, GenEditSC obtains BRISQUE values of approximately $19$--$20$ across the evaluated SNRs and channel types. Its average BRISQUE is $41.47\%$ and $19.33\%$ lower than that of SwinJSCC and SGDJSCC, respectively, over AWGN, and $44.69\%$ and $16.70\%$ lower under Rayleigh fading. 

     {A similar trend is observed on Pascal VOC in Fig.~\ref{fig:pascal_clip_awgn}. GenEditSC obtains an average CLIP-Score that is $6.98\%$ and $8.61\%$ higher than that of SwinJSCC and SGDJSCC, respectively, and average BRISQUE values that are $45.43\%$ and $35.34\%$ lower. Taken together, the DUTS and Pascal VOC results show that GenEditSC is less sensitive to channel degradation over the evaluated operating range; they do not imply invariance to arbitrary channel perturbations.}

\subsubsection{ {Comparison with Generative SemComm}}\label{sec:gen_results}

 {Fig.~\ref{fig:snr_performance} reports CLIP-Score, LPIPS, and BRISQUE versus SNR. At SNR values of 1, 4, 7, and 10~dB, Nam24 obtains CLIP-Scores of $30.41$, $31.90$, $32.15$, and $32.18$; Hosonuma25 obtains $31.52$, $32.45$, $32.56$, and $32.59$; and GenEditSC obtains $33.50$, $33.72$, $33.79$, and $33.80$. Thus, GenEditSC achieves the highest CLIP-Score at each evaluated SNR.}

 {Under the text-transmission setting in Section~V-A, all prompt-based methods use the same UTF-8/LDPC/16-QAM pipeline, while residual decoding errors are retained at low SNR.} Therefore, the CLIP-Score advantage of GenEditSC should be attributed to the complementary visual structural prior and the hybrid communication design,   {which provide complementary semantic information when the recovered prompt is degraded.}

 {GenEditSC also obtains the lowest LPIPS and BRISQUE over the evaluated SNR range. At 1~dB, it retains a measurable margin over Nam24 and Hosonuma25 in both metrics, indicating that the hybrid visual--text condition improves structural fidelity and perceptual naturalness under the tested low-SNR condition.}

 The non-monotonic dataset-average BRISQUE trend of
 Nam24 may result from the interaction between residual textdecoding
 errors and the fixed conditional-generation configuration.
 As SNR increases, the recovered prompt generally
 becomes more reliable, but the dataset-average perceptual
 score need not improve monotonically. GenEditSC exhibits
 lower BRISQUE values and smaller metric changes across the
 tested operating points, which is consistent with the stabilizing
 effect of the transmitted visual prior.

 {Overall, the results show that coordinating a transmitted visual prior with a text-conditioned editing backend improves semantic preservation, perceptual quality, and visual fidelity relative to the evaluated prompt-only and mask-conditioned baselines under the stated settings.}

\begin{figure}[!t]   
	\centering        
	\includegraphics[clip,width=0.84\linewidth]{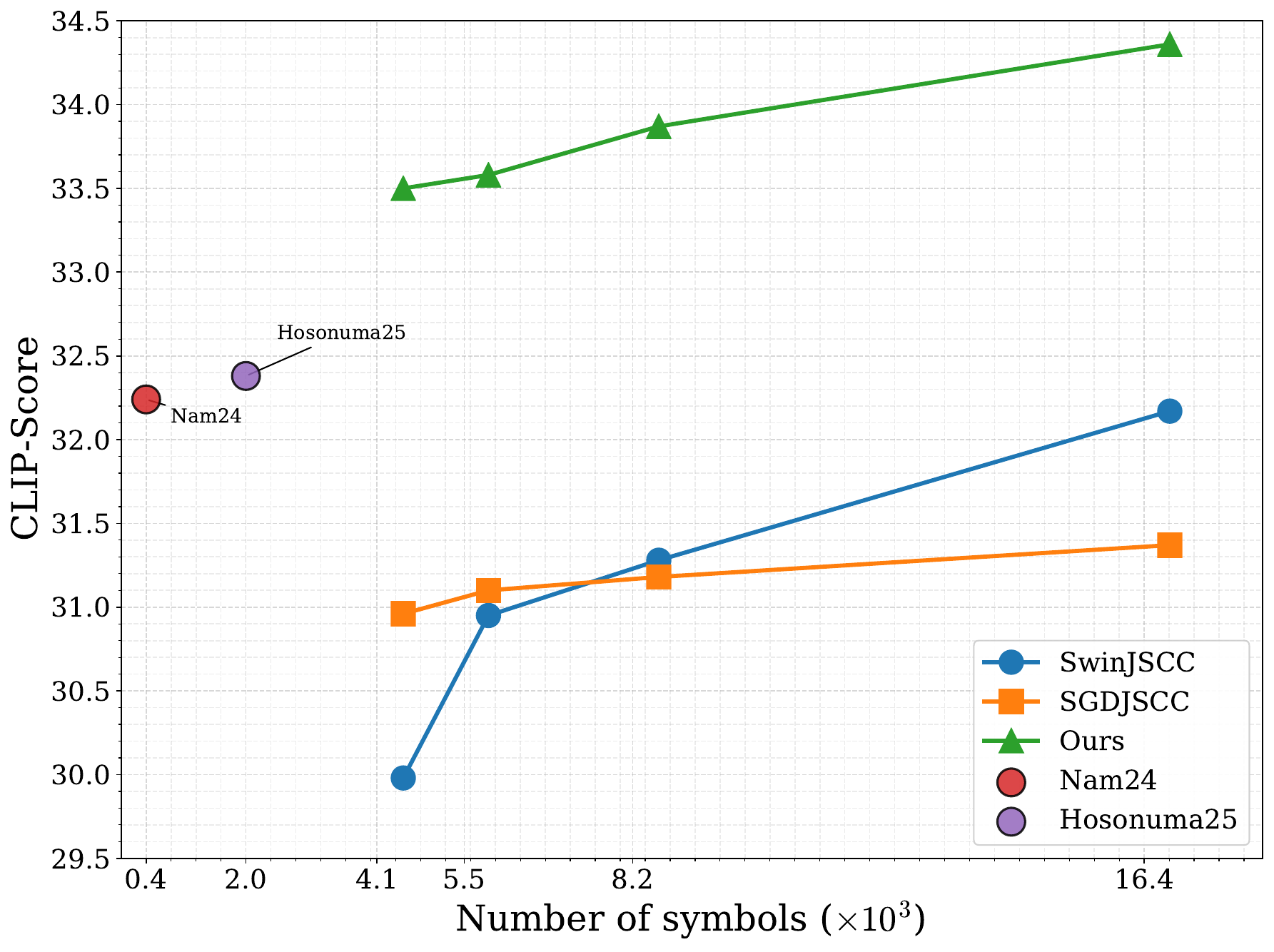}
	\caption{ {\small CLIP-Score versus number of transmitted symbols. Nam24 and Hosonuma25 are fixed-budget baselines and therefore appear as single operating points. SwinJSCC, SGDJSCC, and the proposed method are evaluated under variable symbol budgets.}}\label{bpp}
\end{figure}

    \begin{figure*}[!t]   
        \centering        
        \includegraphics[clip, width=\linewidth]{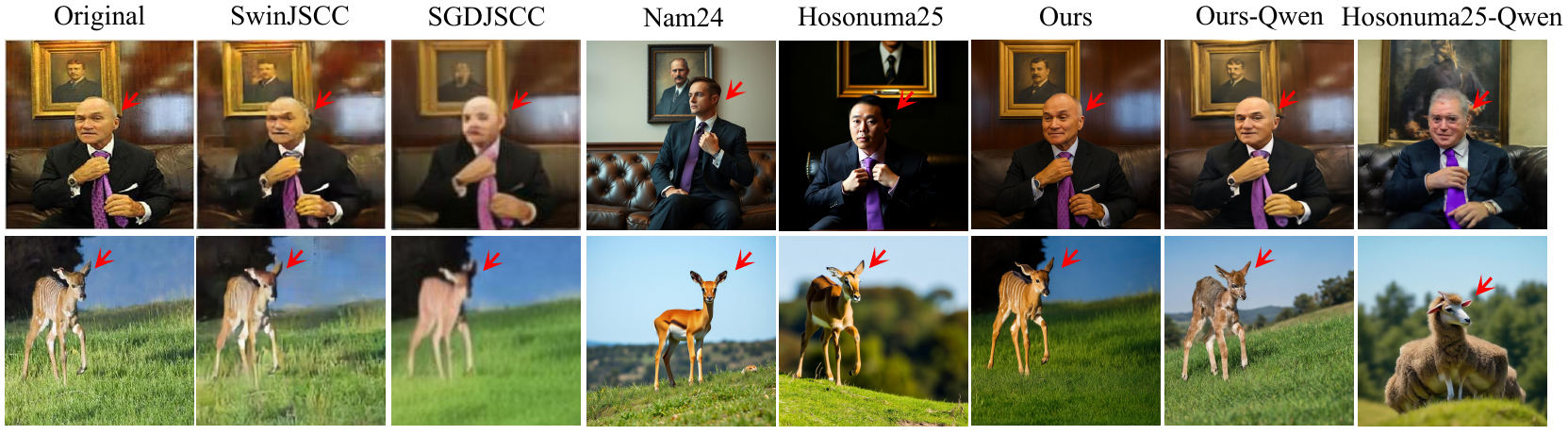}
        \caption{ {\small{Qualitative comparison  at SNR = 4 dB with AWGN channel.}}}
        \label{fig_visual1}
    \end{figure*}
\subsubsection{ {Communication-Overhead Comparison}}\label{sec:overhead_results}

   {In Fig.~\ref{bpp}, we compare GenEditSC with SwinJSCC, SGDJSCC, and representative generative baselines using the communication-overhead accounting defined in Section~\ref{sec:overhead_accounting}.}
  {All experiments are conducted under the same channel condition (\(\mathrm{SNR}=4\) dB).}
   {Across the evaluated compression ratios, the text prompt contributes at most $9.5\%$ of the total symbol count, indicating that semantic text guidance adds only a modest communication cost.}

 {Under the evaluated compression ratios, GenEditSC obtains the highest CLIP-Score among the compared schemes. At approximately $4.5\times10^3$ transmitted symbols, it reaches a CLIP-Score of about $33.5$, whereas SwinJSCC obtains about $30.0$ at a similar symbol budget and remains below GenEditSC when its budget is increased to $1.68\times10^4$ symbols. GenEditSC also remains above SGDJSCC over the tested operating range.}

 {The curves further show a smaller reduction in semantic consistency for GenEditSC in the low-symbol regime. The prompt-only and mask-conditioned baselines operate at small fixed symbol budgets but remain below GenEditSC in CLIP-Score. These results support a better measured trade-off between semantic reconstruction quality and communication overhead under the stated settings. The analysis in Fig.~\ref{bpp} concerns communication overhead only.}
\subsection{{Embedded-Platform Evaluation of the Generative Editing Backend}}
\label{sec:edge_deployment}

	To evaluate receiver-side memory use, per-step computational cost, and reconstruction quality on an embedded platform, we replace the GPT-4o backend used in the main experiments with the Q2\_K-quantized Qwen-Image-Edit model and deploy it on an NVIDIA Jetson AGX Orin 64~GB. Nam24 is not included in this matched-backend evaluation because it uses prompt-only text-to-image generation and therefore requires the Q2\_K-quantized \texttt{Qwen-Image-2512-Q2\_K} model. By contrast, GenEditSC and Hosonuma25 perform image-conditioned editing using \texttt{Qwen-Image-Edit-2511-Q2\_K}.

\begin{table}[!t]
	\centering
	\caption{Receiver-side resource-use comparison.}
	\label{tab:generative_backend_complexity}
	\renewcommand{\arraystretch}{1.12}
	\setlength{\tabcolsep}{5.5pt}
	\begin{tabular}{lcc}
		\toprule
		Method
		& \shortstack{Peak Memory\\on Jetson (GB)}
		& \shortstack{Complexity\\(TFLOPs/Step)} \\
		\midrule
		Hosonuma25 & 21.36 & 245.66 \\
		GenEditSC  & 21.30 & 244.05 \\
		\bottomrule
	\end{tabular}
\end{table}

Table~\ref{tab:generative_backend_complexity} reports the peak
device-memory footprint on the NVIDIA Jetson AGX Orin and the
computational cost of each denoising step. GenEditSC reaches a peak
memory footprint of 21.30~GB. {The measured footprint shows that the quantized generative editing backend fits within the 64-GB memory capacity of the evaluated Jetson platform. GenEditSC and Hosonuma25 also have nearly identical receiver-side complexity. Therefore, the observed reconstruction-performance difference is not explained by differences in the measured backend memory footprint or per-step computational cost.} 

\begin{table}[!t]
	\centering
	\caption{Performance comparison with
		Qwen-Image-Edit on the Jetson platform.}
	\label{tab:qwen_backend_performance}
	\renewcommand{\arraystretch}{1.10}
	\setlength{\tabcolsep}{4.2pt}
	\begin{tabular}{c l c c c}
		\toprule
		SNR
		& Method
		& CLIP-Score $\uparrow$
		& BRISQUE $\downarrow$
		& LPIPS $\downarrow$ \\
		\midrule
		1~dB
		& GenEditSC
		& \textbf{34.32}
		& \textbf{15.13}
		& \textbf{0.399} \\
		&
		Hosonuma25
		& 25.25
		& 23.57
		& 0.699 \\
		\midrule
		4~dB
		& GenEditSC
		& \textbf{34.50}
		& \textbf{15.86}
		& \textbf{0.352} \\
		&
		Hosonuma25
		& 33.69
		& 24.49
		& 0.578 \\
		\midrule
		7~dB
		& GenEditSC
		& \textbf{34.51}
		& \textbf{17.33}
		& \textbf{0.340} \\
		&
		Hosonuma25
		& 34.35
		& 21.73
		& 0.562 \\
		\midrule
		10~dB
		& GenEditSC
		& \textbf{34.52}
		& \textbf{18.32}
		& \textbf{0.337} \\
		&
		Hosonuma25
		& 34.25
		& 21.76
		& 0.561 \\
		\bottomrule
	\end{tabular}
\end{table}

{Table~\ref{tab:qwen_backend_performance} compares GenEditSC and Hosonuma25 using the same Qwen-Image-Edit backend. GenEditSC obtains higher CLIP-Score and lower BRISQUE and LPIPS at every evaluated SNR. The margin is largest at 1~dB: GenEditSC obtains a CLIP-Score of 34.32 and an LPIPS value of 0.399, compared with 25.25 and 0.699 for Hosonuma25, respectively. This result is consistent with the proposed front-end providing a more informative semantic and structural condition to the editor under the evaluated low-SNR setting.}

Averaged over the four operating points, GenEditSC achieves a
CLIP-Score of 34.46, a BRISQUE score of 16.66, and an LPIPS value
of 0.357, whereas Hosonuma25 obtains 31.88, 22.89, and 0.600,
respectively. These results correspond to an 8.08\% improvement in
CLIP-Score and reductions of 27.21\% and 40.50\% in BRISQUE and
LPIPS, respectively. {The results suggest that the semantic-importance-aware JSCC front-end and the spatially anchored reconstructed image contribute to the observed performance advantage.}

   \begin{figure*}[!t]  
        \centering
         \includegraphics[clip,width=\linewidth]{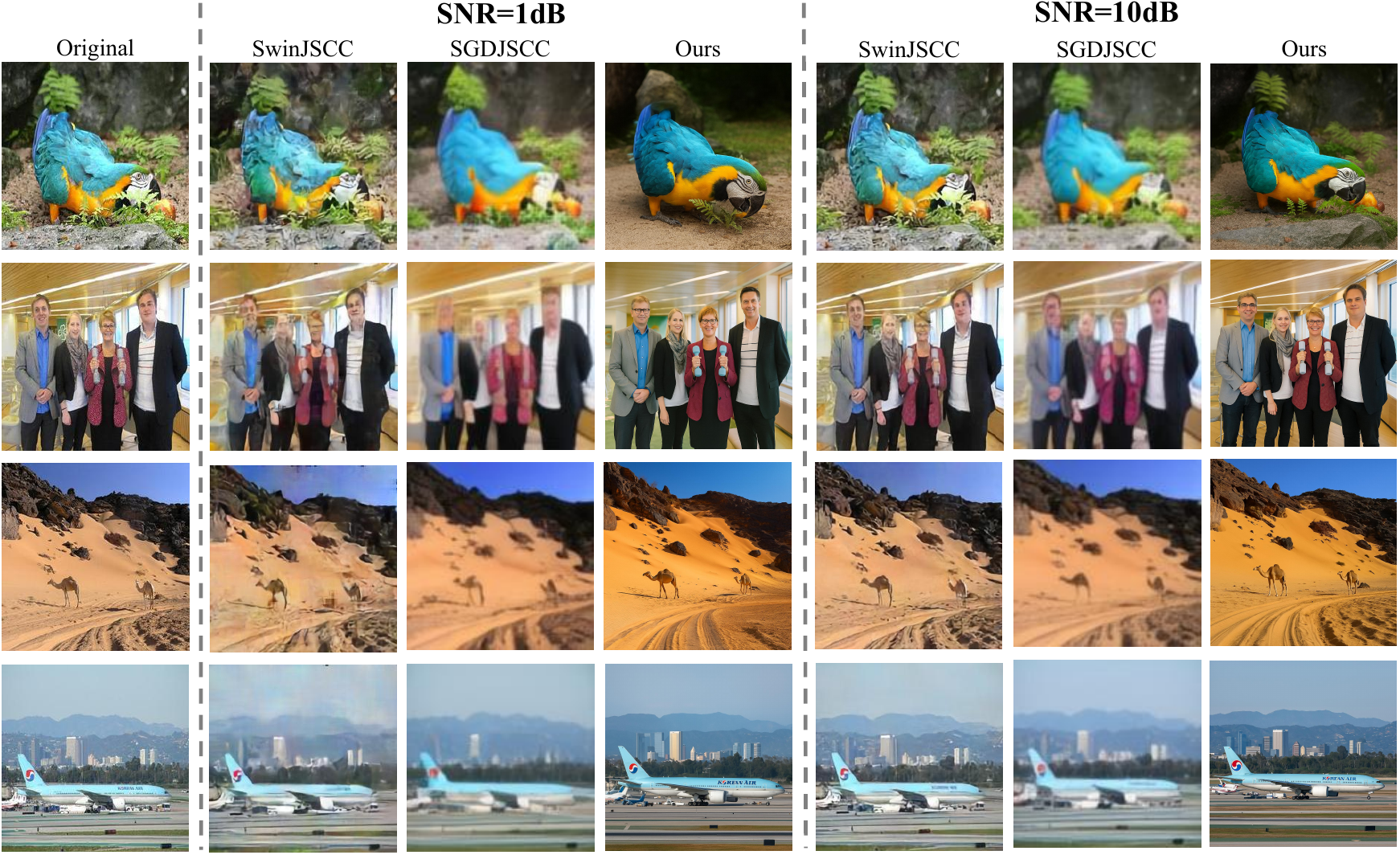}
        \caption{ {\small{Qualitative comparison at SNR values of 1 and 10~dB.}}}
        \label{fig:four_grid}
    \end{figure*}
\subsection{ {Qualitative Results}}\label{sec:qualitative_results} 
\subsubsection{ {Comparison Under a Fixed Channel Condition}}
     {Fig.~\ref{fig_visual1} presents a qualitative comparison between GenEditSC and four representative benchmarks over the AWGN channel at 4~dB. Across the displayed examples, GenEditSC preserves foreground structure and reconstructs background context more consistently than the evaluated baselines. In the first example, it is the only evaluated method that preserves both the subject's facial features and the fine texture of the purple tie while maintaining the wall tone and portrait-frame structure. The observed advantage is consistent with the semantic-importance-aware dual-branch transmission design, which separately represents spatially critical foreground content and less critical background context before generative editing.}

\subsubsection{ {Comparison Across SNRs}}
     {Fig.~\ref{fig:four_grid} presents qualitative comparisons among SwinJSCC, SGDJSCC, and GenEditSC at two channel SNR levels, 1 and 10~dB, across four scenarios: a parrot close-up, an indoor human group, desert camels, and airport aircraft. The examples illustrate how the methods respond to a change in channel quality.}

     {At 1~dB, the discriminative baselines show stronger structural degradation and blur, whereas GenEditSC retains clearer object boundaries and scene layout in the displayed cases. At 10~dB, all methods improve relative to 1~dB, and GenEditSC retains finer details and more natural textures in these examples. The figure supports improved quality at 10~dB relative to 1~dB; it does not by itself establish a proportional relationship between image quality and SNR.}

\subsection{ {Ablation Study}}\label{sec:ablation}

 {We evaluate GenEditSC through component and architecture ablations. All variants use the same main generative editing backend, identical generation settings, and the same evaluation protocol; only the specified communication component or branch architecture is changed. The variants are evaluated using LPIPS, CLIP-Score, and BRISQUE.}
\begin{table}[t]
	\centering
	\caption{ {Component ablation study of GenEditSC.}}
	\label{tab:ablation_components}
	\scalebox{0.86}{
		\begin{tabular}{ccc|ccc}
			\toprule
			\multicolumn{3}{c|}{\textbf{Component}} & \multicolumn{3}{c}{\textbf{Quality Metrics}} \\
			\textbf{Dual} & \textbf{Semantic} & \textbf{Importance} & \textbf{LPIPS}$\downarrow$ & \textbf{CLIP-Score}$\uparrow$ & \textbf{BRISQUE}$\downarrow$ \\
			\textbf{Branch} & \textbf{Objective} & \textbf{Prediction} &  &  &  \\
			\midrule
			&  &  & 0.438 & 32.01 & 21.98 \\
			\checkmark & \checkmark &  & 0.423 & 32.02 & 20.29 \\
			& \checkmark & \checkmark & 0.408 & 32.78 & 20.49 \\
			\multicolumn{3}{c|}{\textbf{Ours}} & \textbf{0.372} & \textbf{33.68} & \textbf{19.72} \\
			\bottomrule
	\end{tabular}}
\end{table}
\begin{table}[t]
\centering
\caption{ {Architecture ablation of the dual-branch design.}}
\label{tab:ablation_arch}
\scalebox{0.9}{
	\begin{tabular}{lccc}
		\toprule
		\textbf{Method} & \textbf{LPIPS}$\downarrow$ & \textbf{CLIP-Score}$\uparrow$ & \textbf{BRISQUE}$\downarrow$ \\
		\midrule
		Swin branch & \underline{0.397} & \underline{33.02} & \underline{20.04} \\
		CBAM branch & 0.408 & 32.78 & 20.49 \\
		Ours & \textbf{0.372} & \textbf{33.68} & \textbf{19.72} \\
		\bottomrule
\end{tabular}}
\end{table}
\subsubsection{ {Component Ablation}}

 {Table~\ref{tab:ablation_components}  reports the ablation results of the three key designs, namely the dual-branch transmission architecture, the semantic importance prediction module, and the semantic-importance-aware objective. Starting from the basic variant, introducing either semantic-aware decomposition or semantic guidance leads to consistent improvement. The full model achieves the best performance on all three metrics, with LPIPS = 0.372, CLIP-Score = 33.68, and BRISQUE = 19.72. These results indicate that the gain does not come from any single off-the-shelf component, but from the coupled design of semantic importance prediction, dual-branch transmission, and semantic-aware optimization.}

\subsubsection{ {Architecture Ablation}}

 {To further justify the proposed asymmetric dual-branch design, we compare it with two homogeneous alternatives: \emph{Swin branch}, where both branches use the same structure-oriented encoder, and \emph{CBAM branch}, where both branches use the same attention-oriented refinement mechanism. As shown in Table~\ref{tab:ablation_arch} , both alternatives are inferior to the proposed design. Using Swin in both branches tends to over-preserve redundant background content, while using CBAM in both branches weakens the distinction between important and redundant regions. In contrast, the proposed asymmetric design achieves the best trade-off between semantic fidelity and perceptual quality, confirming that its advantage comes from differentiated functional roles for foreground and background transmission rather than from increased model complexity alone.}

\section{Conclusion}\label{conclusion}
     {This paper proposed GenEditSC, a semantic-importance-aware image-transmission framework that combines a trainable JSCC communication front-end with a fixed and replaceable multimodal image editing backend. The JSCC stage prioritizes semantically important regions to preserve scene structure under bandwidth constraints, and the editing stage refines missing or degraded details using the reconstructed image and a transmitted text prompt. Experiments on DUTS and Pascal VOC show that GenEditSC outperforms the evaluated discriminative and generative baselines in semantic fidelity, perceptual quality, and communication efficiency across the considered channel conditions. In addition, deploying a 2-bit-quantized Qwen-Image backend on an NVIDIA Jetson AGX Orin 64~GB validates the edge deployment feasibility. These results indicate that the proposed communication architecture is compatible with different generative editing implementations.}

      {This study focuses on semantic image transmission, while the decoupled design of the trainable JSCC front-end and the fixed, replaceable generative editing backend provides a general and extensible basis beyond the particular backend implementation and image-transmission setting considered here.}
      {Potential extensions include video transmission and user-controllable editing mechanisms tailored to diverse downstream applications.}

\section*{Appendix}\label{app:score_map}

    This appendix provides the detailed implementation for generating the Semantic Importance Score Map $\bm{S}$ from the enhanced text embedding $\tilde{\bm{e}}_\mathrm{t}$ and the visual embedding $\bm{e}_\mathrm{v}$, as referenced in Section IV-A.

    By concatenating $\bm{g}$ and $\bm{e}_\mathrm{v}$ along the spatial dimension, we directly form the visual context $\bm{e} \in \mathbb{R}^{(1+H \times W) \times C}$:
	\begin{equation}
		\bm{e}_\text{context} = \left[ \bm{g}; \bm{e}_\mathrm{v} \right],
	\end{equation}
	where $\bm{g}$ is broadcasted to \( \mathbb{R}^{1 \times C} \) and $\bm{e}_\mathrm{v}$ is implicitly interpreted as \( \mathbb{R}^{HW \times C} \).

	To improve pixel-level semantic alignment between text and image, we propose a CLIP-based cross-modal context matching module that enhances the original text embedding with visual cues. 
	The visual context $\bm{e}_\text{context}$ is then fed into a Transformer-based cross-modal decoder module $\mathcal{D}_\text{ctx}$ together with the text embedding as query:
	\begin{equation}
	\Delta \bm{e}_\mathrm{t} = \mathcal{D}_\text{ctx}(\bm{e}_\mathrm{t}, \bm{e}_\text{context}) \in \mathbb{R}^{C}.
	\end{equation}
	The enhanced text embedding $\tilde{\bm{e}}_t\in \mathbb{R}^{C}$ is given by:
	\begin{equation}
	\tilde{\bm{e}}_\mathrm{t} = \bm{e}_\mathrm{t} + \gamma \cdot \Delta \bm{e}_\mathrm{t} ,
	\end{equation}
	where $\gamma$ is a learnable scalar parameter to control the fusion strength. This step effectively aligns the text feature to
	the image semantics by incorporating local visual context.

    Finally, the pixel-wise semantic importance score $\bm{S}_{i,j}$ at each spatial location $(i, j)$ is computed as the scaled dot-product similarity:
    	\begin{equation}
    	{\bm{S}}_{i,j} = \frac{1}{\tau} \sum_{c=1}^{C} \tilde{\bm{e}}_\mathrm{t}^{(c)} \cdot \bm{e}_\mathrm{v}^{(c, i, j)}, \quad \forall i \in [1, H],\ j \in [1, W],
            \label{eq:semantic_score}
    	\end{equation}
    	where $\tau$ is a learnable temperature parameter that controls the sharpness of the similarity scores.

\bibliography{ref}    
\bibliographystyle{ieeetr}

\end{document}